\documentclass[a4paper,prd,twocolumn,showpacs,superscriptaddress,floatfix,nofootinbib]{revtex4-2}
\usepackage{graphicx}
\usepackage{amsmath}
\usepackage{amssymb}
\usepackage{enumitem}
\usepackage{xcolor}
\usepackage{xspace}
\usepackage{hyperref}
\usepackage{times}
\hypersetup{
    colorlinks=true, linkcolor=blue, citecolor=blue,
    filecolor=magenta, urlcolor=cyan}

\newcommand{\eg}{e.g.,\xspace}
\newcommand{\ie}{i.e.,\xspace}
\newcommand{\rns}{\texttt{RNS}\xspace}
\newcommand{\Msun}{\,M_{\odot}}
\newcommand{\orcid}[1]{\href{https://orcid.org/#1}{
\includegraphics[width=10pt]{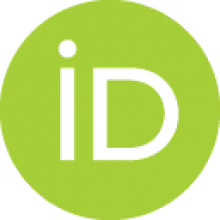}}}

\begin{document}

\title{Energy budget for magnetic field amplification in long-lived neutron star merger remnants}

\author{Jin-Liang Jiang\:\orcid{0000-0002-9078-7825}}
\affiliation{Department of Astronomy, School of Physics and Astronomy,
  Key Laboratory of Astroparticle Physics of Yunnan Province,
  Yunnan University, Kunming 650091, People's Republic of China}

\author{Ming-Zhe Han\:\orcid{0000-0001-9034-0866}}
\email{hanmz@pmo.ac.cn}
\affiliation{Key Laboratory of Dark Matter and Space Astronomy, Purple Mountain Observatory, Chinese Academy of Sciences, Nanjing, 210033, People's Republic of China}
\affiliation{Max-Planck-Institut f\"ur Gravitationsphysik (Albert-Einstein-Institut), Am M\"uhlenberg 1, D-14476 Potsdam-Golm, Germany}

\date{\today}

\begin{abstract}
The upper limit of magnetic field amplification in binary neutron star (BNS) mergers remains unknown.
Here we address the limit set by the differential rotation of the remnant.
Three processes are widely known to convert the energy stored in differential rotation into magnetic energy and drive the merger remnant into uniform rotation: winding, the magnetorotational instability (MRI), and the Tayler--Spruit instability (TSI).
Because these processes are effectively activated only in a differentially rotating remnant, the energy difference between the differentially rotating configuration and its uniformly rotating counterpart provides an upper limit on the electromagnetic (EM) energy that they can generate.
Using simulation results for different equations of state, and taking into account different angular momentum loss mechanisms for the post-merger remnant, we find an energy reservoir of $(3$--$7)\times10^{52}~\mathrm{erg}$ for spinning the differentially rotating remnant down to uniform rotation, which sets an upper limit on the magnetic field of $(2$--$4)\times10^{17}~\mathrm{G}$ that can be amplified by differential rotation and leaves ample room for simulations of magnetic field amplification in the BNS merger remnant.
The non-collapsing remnants also retain a rotation frequency of $1$--$1.4~\mathrm{kHz}$ and an energy reservoir of uniform rotation of $(4$--$12)\times10^{52}~\mathrm{erg}$, comparable to or up to a few times larger than the braking energy of the differential rotation; this energy is to be injected into the surrounding environment and can power luminous EM transients.
These results can be tested by future ultra-high-resolution numerical simulations.
\end{abstract}
\maketitle

\section{Introduction}
\label{sec:intro}

The first gravitational-wave (GW) detection of a binary neutron star (BNS) merger, GW170817, and its electromagnetic (EM) counterpart~\cite{Abbott2017_GW170817_find, Abbott2017_GW170817_multimessenger, Abbott2018_GW170817_property} have greatly advanced our understanding of neutron star (NS) properties and stimulated a wide range of BNS merger simulations.
On the observation side, the counterparts of GW170817 span multiband EM emission from the gamma-ray burst (GRB)~\cite{FermiLat2017, Goldstein2017}, multiband afterglow emission~\cite{Hallinan2017, Troja2017, Margutti2018, Mooley2018}, and kilonova~\cite{Chornock2017, Drout2017, Nicholl2017, Pian2017, Smartt2017, Soares-Santos2017}, as well as afterglow signals lasting from months to years~\cite{Troja2019, Hajela2022}.

On the theoretical side, previous general-relativistic magnetohydrodynamic (GRMHD) simulations and theoretical calculations have targeted various aspects of the observational signatures generated by BNS mergers.
These include GW emission~\cite{Baiotti2016, Paschalidis2016a, Radice2020, Huang2026a}, multiwavelength EM signals from magnetized flares~\cite{Most2023a, Musolino2024b, Jiang2025}, extended emission~\cite{Musolino2024a}, short GRBs from relativistic jets~\cite{Giacomazzo2011, Rezzolla2011, Baiotti2016, Ciolfi2020, Kiuchi2023, Gottlieb2024, Rainho2025}, afterglow~\cite{WangH2024}, kilonovae produced by the decay of \(r\)-process nuclei in the ejecta~\cite{Li1998, Radice2016, Metzger2017, Bovard2017, Zhu2021b, Combi2022, Fujibayashi2023, Kawaguchi2023, Cheong2024, Ng2024}, and the production of heavy elements~\cite{Metzger2010, Bovard2017, Papenfort2018, Combi2022, Fujibayashi2023, Kawaguchi2023}.

Magnetic fields are vital in the evolution of a BNS system.
Many studies have found that the merger remnant can reach magnetar-level field strengths~\cite[see, \eg][]{Duncan1992, Giacomazzo2014, Kiuchi2017, Palenzuela2022, Combi2023, Miret2025, Gutierrez2026, Kiuchi2026a}.
During the inspiral, the magnetic field plays a minor role in the orbital dynamics~\cite{Giacomazzo2009, Zhu2020}, but it becomes crucial during and after the merger, particularly for the jets that power GRBs.
Various magnetic field amplification processes operate throughout the merger, as revealed by previous studies (see also~\cite{Kiuchi2024_review} for a comprehensive review).
These include the Kelvin--Helmholtz instability (KHI; see~\cite{Rasio99, Price06, Baiotti08, Giacomazzo2014, Kiuchi2015, Chabanov2022, Kiuchi2023, Bamber2024a, Cook2025, Neuweiler2025, Ng2026a, Gutierrez2026, Kiuchi2026a, Kiuchi2026b}, also see shearing-box simulations~\cite{Obergaulinger10, Zrake2013b}), winding~\cite{Kluzniak1997, Shapiro2000, Liu2003, Duez2006, Shibata2021a, Combi2023, Miret2023, Gutierrez2025}, the magnetorotational instability (MRI)~\cite{Velikhov1959, Chandrasekhar1960, Balbus1991, Rezzolla2011, Siegel2013, Moesta2020, Kiuchi2023, Bamber2024b, Celora2025, Salze2025, Kiuchi2026a, Kiuchi2026b}, and the Tayler--Spruit instability (TSI)~\cite{Tayler1973, Spruit2001, Salze2024, Ng2026b}, among others.
Strong magnetic fields resulting from these amplification processes can eventually lead to wind-driven outflows~\cite{Siegel2014} and allow the magnetic field to break out of the stellar interior~\cite{Parker1966, Kluzniak1997, Dai2006, Kiuchi2011, Most2023a, Musolino2024b, Jiang2025, Fields2025, Wen2026}, causing X-ray flares, modifying the properties of kilonovae~\cite{Feng2026}, and even powering the GRB itself~\cite{Kluzniak1997, Moesta2020, Desai2026, Kiuchi2026b}.

However, the maximum magnetic field strength achievable by the remnant remains an open question.
This question is challenging because of the vast range of scales involved, as KHI, MRI, and TSI amplify the field even at wavelengths far below the resolution of current merger simulations~\cite[see, \eg][]{Kiuchi2015, Siegel2013, Salze2024}.
Consequently, tracking the amplification to saturation requires resolutions well beyond current computational capabilities.
Although recent high-resolution direct numerical simulations~\cite{Gutierrez2026, Kiuchi2026a} and large-eddy simulations~\cite{Palenzuela2022, Miret2025} report magnetic field saturation, these findings still require independent verification by other groups at higher resolutions and over a broader parameter space.

In this work, we study EM energy amplification using a straightforward approach that circumvents the need for extremely high-resolution simulations, namely an energy budget analysis based on numerical simulations that employ realistic equations of state (EOSs).
We therefore ask: if the merger remnant does not collapse for a long time, what is the maximum EM energy that it can reach in principle?
The underlying rationale is that, among the four well-known magnetic amplification processes discussed above, the latter three (\ie winding, MRI, and TSI) rely on the differential rotation of the post-merger remnant.
These processes convert kinetic energy into magnetic and thermal energy while transporting angular momentum inward or outward, and ultimately drive the remnant toward a uniformly rotating configuration (unless it collapses during the process).
Consequently, the upper limit of magnetic field amplification is set by the energy difference between the differentially rotating configuration and the corresponding uniformly rotating one.
This upper limit should guide theoretical and semi-analytical analyses of the energy injection from the merger-formed magnetar into the GRB, and can be verified by future, more accurate simulations.

The paper is structured as follows.
Section~\ref{sec:methods} describes how the uniformly and differentially rotating configurations are computed, how the braking energy of the differential rotation and that of the uniform rotation are defined, and how the energy budget is calculated under different possible angular momentum loss regimes.
Section~\ref{sec:results} presents the properties of the long-term remnant in the simulation and of the final remnant after angular momentum loss, along with the observational implications.
Section~\ref{sec:discussion} summarizes our findings and their limitations.
Throughout this work we use units with $c=G=1$ unless otherwise specified.

\section{Methods}
\label{sec:methods}

We aim to establish the connection between the differentially rotating configuration and the uniformly rotating configuration in order to determine the upper limit of magnetic field amplification due to differential rotation.
We first need uniformly rotating configurations, which are constructed with the \rns code~\cite{Stergioulas1995}, and then differentially rotating configurations, which are taken directly from the numerical-relativity simulations of BNS mergers in Ref.~\cite{Han2025}.
The following subsections describe how the energy budgets in both differentially and uniformly rotating configurations are calculated and how the different angular momentum loss channels affect them.

\subsection{Uniformly rotating configurations}

For a stationary, axisymmetric perfect fluid in circular motion, the line element of spacetime reads (in quasi-isotropic coordinates)
\begin{equation}
  \label{eq:metric}
  \begin{aligned}
  g_{\alpha\beta}\,\mathrm{d}x^\alpha\,\mathrm{d}x^\beta = &\; -N^2\mathrm{d}t^2 + A^2(\mathrm{d}r^2+r^2\mathrm{d}\theta^2) \\
  & + B^2 r^2\sin^2\theta (\mathrm{d}\varphi -\omega\mathrm{d}t)^2.
  \end{aligned}
\end{equation}
Here $r$ is the coordinate radius, the equatorial circumferential radius is $R\equiv\sqrt{g_{\varphi\varphi}}|_{\theta=\pi/2}=Br$, and $\varpi\equiv\sqrt{g_{\varphi\varphi}}=Br\sin\theta$ is the circumferential (cylindrical) radius.
If we further require that $j\equiv u^t u_{\varphi}$ be a function of the angular velocity $\Omega\equiv u^\varphi/u^t$ only (the rotation profile), \ie
\begin{equation}
  \label{eq:rot_prof}
  j = j(\Omega),
\end{equation}
then a first integral can be found for the fluid motion~\cite{Gourgoulhon2010}. The first integral reads:
\begin{equation}
  \label{eq:first_int}
  H+\ln N-\ln{\Gamma}+\int_{\Omega_\mathrm{c}}^{\Omega} j(\Omega^\prime)\, \mathrm{d}\Omega^\prime = \mathrm{const},
\end{equation}
where $H\equiv\ln[(e+p)/\rho]$ is the logarithmic specific enthalpy; $e$, $p$, and $\rho$ are the energy density, pressure, and rest-mass density; $\Gamma$ is the Lorentz factor of the fluid with respect to the zero-angular-momentum observer; and $\Omega_\mathrm{c}$ is the central angular velocity.

This first integral is used in the \rns code to construct uniformly or differentially rotating NS configurations with the self-consistent field method~\cite{Friedman1986, KEH-I, KEH-II, CST94, Stergioulas1995, Nozawa1998, Friedman-Nikolaos2013}. For the uniformly rotating configurations considered in this work, the fourth term on the left-hand side of Eq.~\eqref{eq:first_int} vanishes.

Regarding the code setup in \rns, we choose a grid of $(\texttt{MDIV}\times \texttt{SDIV}=129 \times 257)$ zones. An axis-ratio step of $0.005$ is used when stepping iteratively from the non-rotating model to the mass-shedding limit at fixed central density for each EOS. Both parameters were further verified with higher resolution and found to be sufficient.

\subsection{Differentially rotating configurations}

We characterize differentially rotating merger remnants directly from 3D numerical-relativity simulations, rather than by prescribing an analytic rotation law.
The underlying BNS simulations were performed with the \texttt{SACRA-MPI} code~\cite{Yamamoto2008}.

\begin{figure*}[!t]
  \centering
  \includegraphics[width=0.4654\textwidth]{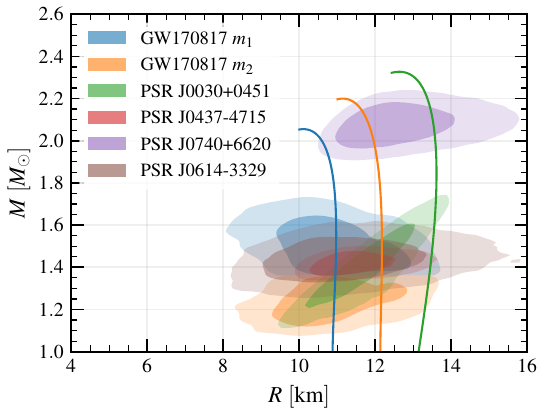}
  \includegraphics[trim=0bp 0.34277bp 0bp 0bp,width=0.48\textwidth]{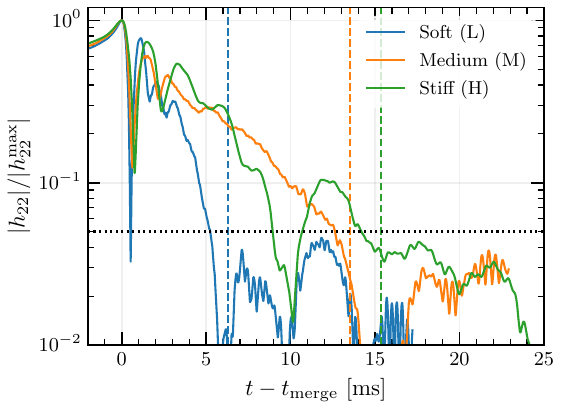}
  \caption{Left panel: $M$--$R$ curves for the three selected cold EOSs: soft (L, blue), medium (M, orange), and stiff (H, green). The darker and lighter shaded regions indicate the 68\% and 95\% credible regions, respectively, for the observational $M$--$R$ constraints from the two components of GW170817 and the pulsars PSR J0030+0451, PSR J0437$-$4715, PSR J0740+6620, and PSR J0614$-$3329. Right panel: time evolution of the normalized amplitude of the dominant $(\ell,m)=(2,2)$ GW strain mode for representative BNS merger simulations with different EOSs but the same thermal index $\Gamma_\mathrm{th}=2.0$ and total mass $M_\mathrm{tot}=2.7\Msun$. The vertical dashed lines mark the selected time slices for each model used in the subsequent analysis, and the horizontal black dotted line is the $1/20$ threshold.}
  \label{fig:eos}
\end{figure*}

Three typical cold NS EOSs that satisfy the current observational and theoretical constraints are chosen for the simulations.
A detailed description of the EOS construction and inference process can be found in Refs.~\cite{Han2021, Han2022}.
For completeness, we list the macroscopic properties of the cold NS EOSs in Table~\ref{tab:EOSs}, and show their mass--radius ($M$--$R$) curves along with the multimessenger constraints in the left panel of Fig.~\ref{fig:eos}.
These are the mass--radius constraints~\cite{Abbott2018_GW170817_EOS} inferred from the mass and tidal deformability measurements from GW170817~\cite{Abbott2017_GW170817_find} and the simultaneous measurements of mass and radius from the NICER observations, namely PSR J0030+0451~\cite{Vinciguerra2024}, PSR J0740+6620~\cite{Salmi2024}, PSR J0437$-$4715~\cite{Choudhury2024}, and PSR J0614$-$3329~\cite{Miller2026}.
Finally, the thermal effect is modeled through the hybrid EOS prescription~\cite{Janka1993}
\begin{equation}
  \begin{aligned}
  \varepsilon &= \varepsilon_\mathrm{cold}(\rho)+\varepsilon_\mathrm{th},\\
  p &= p_\mathrm{cold}(\rho)+(\Gamma_\mathrm{th}-1)\rho\varepsilon_\mathrm{th},
  \end{aligned}
\end{equation}
where $\varepsilon=e/\rho - 1$ is the specific internal
energy, and $\Gamma_\mathrm{th}$ is a constant thermal index for each simulation.
This prescription permits a controlled exploration of thermal support while keeping the cold EOS fixed.

To obtain a remnant that can be approximately compared with a stationary, axisymmetric rotating star, we select a 3D snapshot after the strong post-merger quadrupolar GW oscillations have subsided. We define $t_\mathrm{merge}$ as the time of the maximum $(\ell,m)=(2,2)$ GW strain amplitude ($h_{22}$) and take $t_\mathrm{s}$ to be the first available 3D snapshot after $t_\mathrm{cross}$, the earliest time after which the averaged GW amplitude stays below $1/20$ of its peak value for $2$~ms, \ie
\begin{align}
  \langle|h_{22}|\rangle(t)
  &=\frac{1}{\Delta}\int_{t-\Delta/2}^{t+\Delta/2}|h_{22}(t')|\,\mathrm{d}t',
  \quad \Delta=2~\mathrm{ms},\\
  t_\mathrm{cross}
  &=\min\Bigl\{t:\ \langle|h_{22}|\rangle(t')<0.05\,|h^\mathrm{max}_{22}|,\nonumber\\
  &\qquad\qquad \forall\,t'\in[t,\,t+2~\mathrm{ms}]\Bigr\}.
  \label{eq:ts}
\end{align}
The final selected time lies between $6.3$ and $28.1~\mathrm{ms}$ (second column of Table~\ref{tab:transport}).
We show the GW signal of three representative models with thermal index $\Gamma_\mathrm{th}=2$ and total mass $M_\mathrm{tot}=2.7\Msun$ in the right panel of Fig.~\ref{fig:eos}. As shown in the figure, a remnant with a stiffer EOS shows a slower decay of the post-merger GW signal, since the compactness strongly affects the dynamical timescale over which the fluid readjusts.
Since $h_{22}$ probes primarily the non-axisymmetric $m=2$ deformation, the decay of its dominant part motivates a quasi-axisymmetric and nearly stationary description, which allows a well-defined gravitational mass and angular momentum enclosed by a surface of constant cut-off density, and a direct comparison of the Arnowitt--Deser--Misner (ADM) mass and the Komar mass in a volume integral form (see Appendix~\ref{app:global_quantities}).

The simulations used here are the highest-resolution runs that do not collapse within $25~\mathrm{ms}$ after merger in Ref.~\cite{Han2025}, and only equal-mass binaries are considered here, \ie $m_\mathrm{i}=M_\mathrm{tot}/2=\{1.35, 1.40, 1.45\} \Msun$; the code setup and implementation details can be found there. For comparison with the cold uniformly rotating \rns equilibria, we evaluate the baryon mass, gravitational mass, angular momentum, and thermal energy in the dense region $\rho\geq\rho_\mathrm{cut}=10^{11}\,\mathrm{g\,cm^{-3}}$, using the Komar-based definitions described in Appendix~\ref{app:global_quantities}.

\begin{table}[htbp]
  \centering
  \caption{Properties of the EOSs used in this work. From left to right, the maximum gravitational mass of the non-rotating cold sequence, the Tolman--Oppenheimer--Volkoff (TOV) mass $M_\mathrm{TOV}$; that of the mass-shedding sequence (the Keplerian rotating models), $M_\mathrm{kep}^{\max}$; the circumferential radii at the equator of $1.4\Msun$ and $2.0\Msun$ non-rotating NSs, $R_{1.4}$ and $R_{2.0}$; and the dimensionless tidal deformability of a $1.4\Msun$ NS, $\Lambda_{1.4}$.}
  \label{tab:EOSs}
  \begingroup
  \setlength{\tabcolsep}{8.5pt}
  \renewcommand{\arraystretch}{1.05}
  \small
  \begin{tabular}{l cccc c}
  \hline
  \hline
  EOS & $M_\mathrm{TOV}$ & $M_\mathrm{kep}^{\max}$ & $R_{1.4}$ & $R_{2.0}$ & $\Lambda_{1.4}$ \\
  & $[M_\odot]$ & $[M_\odot]$ & $[\mathrm{km}]$ & $[\mathrm{km}]$ & \\
  \hline
  Soft   & $2.05$ & $2.48$ & $10.96$ & $10.61$ & $196$ \\
  Medium & $2.20$ & $2.66$ & $12.18$ & $11.96$ & $397$ \\
  Stiff  & $2.32$ & $2.86$ & $13.45$ & $13.58$ & $800$ \\
  \hline
  \hline
  \end{tabular}
  \endgroup
\end{table}

\subsection{Energy budget of differential rotation}
\label{sec:budget}

At first sight, the energy budget for magnetic field amplification is the difference of kinetic energy between the differentially rotating and the uniformly rotating configuration.
However, the energy balance shows that the true budget lies in the difference of gravitational mass between the two configurations.
The main reason we cannot directly use kinetic energy is that the density profile changes while the remnant spins down, so gravitational binding energy can be released into kinetic energy.
The gravitational binding energy is generally several to more than ten times the rotational kinetic energy in uniformly rotating supramassive NSs, thus its change cannot be neglected.
On the other hand, the gravitational mass contains the rest mass, the kinetic, internal, and magnetic energies, and the gravitational potential energy from the Newtonian point of view.
Consequently, comparing the gravitational mass difference between two configurations with the same baryon mass and the same total angular momentum isolates the energy budget for the three main magnetic field amplification channels in the post-merger phase (\ie winding, MRI, and TSI), which rely heavily on the differentially rotating nature of the remnant in amplifying the magnetic field.

When comparing two isolated configurations with the same baryon mass \(M_\mathrm{b}\) and the same total angular momentum \(J\), we let \(M_\mathrm{g}^\mathrm{D}(M_\mathrm{b}, J)\) and \(M_\mathrm{g}^\mathrm{U}(M_\mathrm{b}, J)\) be the gravitational masses of the initial differentially rotating configuration and of the unmagnetized uniformly rotating cold reference equilibrium, respectively.
In the idealized case, both \(M_\mathrm{b}\) and \(J\) are conserved.
The braking energy available from the differential rotation is then
\begin{equation}
  E^\mathrm{D}_\mathrm{brake} = M_\mathrm{g}^\mathrm{D}(M_\mathrm{b}, J) - M_\mathrm{g}^\mathrm{U}(M_\mathrm{b}, J).
  \label{eq:free-energy-ideal}
\end{equation}

However, in a realistic evolution the remnant is not isolated, and three main channels remove energy from it: GWs (\(E_\mathrm{GW}\)), neutrinos (\(E_\nu\)), and mass ejection (\(E_\mathrm{eje}\)).
The final gravitational mass is therefore reduced by the total energy loss, and the energy balance reads
\begin{equation}
  M_\mathrm{g}^\mathrm{D}(M_\mathrm{b}^\mathrm{D}, J_\mathrm{D}) = M_\mathrm{g}^\mathrm{U}(M_\mathrm{b}^\mathrm{U}, J_\mathrm{U}) + E_\mathrm{mag}+E_\mathrm{loss}+E_\mathrm{res},
  \label{eq:mass-balance}
\end{equation}
where $E_\mathrm{res}$ represents the residual energy that is neither transferred into magnetic energy nor contained in the cold equilibrium sequences. The total energy loss is
\begin{equation}
  E_\mathrm{loss} = E_\mathrm{GW} + E_\nu + E_\mathrm{eje},
  \label{eq:Delta-M-loss}
\end{equation}
and the ejecta carry the energy
\begin{equation}
  E_\mathrm{eje} = \Gamma_\mathrm{eje} M_\mathrm{eje},
  \label{eq:e_ejecta}
\end{equation}
with $\Gamma_\mathrm{eje}$ the Lorentz factor of the ejecta.
Correspondingly, the baryon masses are related by
\begin{equation}
  \label{eq:Mb-loss}
  M_\mathrm{b}^\mathrm{D}=M_\mathrm{b}^\mathrm{U}+M_\mathrm{eje}.
\end{equation}

Angular momentum leaves mainly through the same three channels, so the angular momentum of the final reference state is reduced accordingly.
Denoting the total angular momentum loss by \(J_\mathrm{loss}\), we have
\begin{equation}
  J_\mathrm{D} = J_\mathrm{U} + J_\mathrm{loss},
  \label{eq:J-balance}
\end{equation}
where
\begin{equation}
  J_\mathrm{loss} = J_\mathrm{GW} + J_\nu + J_\mathrm{eje}.
  \label{eq:J-loss}
\end{equation}
Here \(J_\mathrm{GW}\), \(J_\nu\), and \(J_\mathrm{eje}\) are the angular momenta carried away by GWs, neutrinos, and ejected mass, respectively.
Note that internal processes such as thermalization, oscillations, and turbulence redistribute angular momentum within the star, but they do not change the total angular momentum unless they feed one of these loss channels.
Thus the final uniformly rotating reference state is the initial remnant with the baryon mass reduced by \(M_\mathrm{eje}\) and the angular momentum reduced by \(J_\mathrm{loss}\) (for the details of mass and angular momentum loss, see~Sec.~\ref{sec:loss}).

Bearing all the above arguments in mind, the braking energy of the differential rotation reads
\begin{equation}
  E^\mathrm{D}_\mathrm{brake} = M_\mathrm{g}^\mathrm{D}(M_\mathrm{b}^\mathrm{D}, J_\mathrm{D}) - M_\mathrm{g}^\mathrm{U}(M_\mathrm{b}^\mathrm{U}, J_\mathrm{U}) - E_\mathrm{loss},
  \label{eq:free-energy-final}
\end{equation}
with the baryon-mass and angular-momentum differences defined by Eqs.~\eqref{eq:Mb-loss} and~\eqref{eq:J-balance}, respectively.
This differential braking energy bounds the magnetic energy that can be generated through differential rotation, \ie
\begin{equation}
  E_\mathrm{mag} \leq E^\mathrm{D}_\mathrm{brake},
  \label{eq:EB-ideal}
\end{equation}
where the equality holds only if all of the differential braking energy is converted into magnetic energy, \ie $E_\mathrm{res}=0$.

\subsection{Energy budget of uniform rotation}
\label{sec:uniform}

Once the differential rotation is lost, the remnant is a uniformly rotating star, and it still holds a reservoir of energy~\cite{Metzger2015, Beniamini2021, Tang2025a}.
In this section we discuss how much energy the remnant can still release if it spins down to the smallest angular momentum configuration that its baryon mass allows.
We measure this reservoir in the same way as $E^\mathrm{D}_\mathrm{brake}$, \ie by the difference of gravitational masses at fixed baryon mass.
Let $M_\mathrm{g}^\mathrm{U}(f)$ be the gravitational mass $M_\mathrm{g}^\mathrm{U}$ of Sec.~\ref{sec:budget} (\ie the end state described in~Sec.~\ref{sec:loss}), now labeled by the rotation frequency $f$ of the uniformly rotating equilibrium at fixed baryon mass $M_\mathrm{b}$, and $M_\mathrm{g}^\mathrm{ref}$ that of the reference configuration of the same baryon mass, which rotates at $f^\mathrm{ref}$.
Which configuration is the reference depends on $M_\mathrm{b}$.
Below the maximum baryon mass of the non-rotating sequence, $M_\mathrm{b}^\mathrm{TOV}$, a non-rotating star exists and is the reference, $f^\mathrm{ref}=0$.
Above it no non-rotating equilibrium exists and only rotation can support that mass, so the reference is the configuration at the turning-point line of the sequence~\cite{Metzger2015, Beniamini2021, Tang2025a}, the locus of the turning points of its constant angular momentum sequences; beyond the line the configuration is secularly unstable and may collapse~\cite{Friedman1988}.
The braking energy of the uniform rotation is therefore
\begin{equation}
  E^\mathrm{U}_\mathrm{brake} = M_\mathrm{g}^\mathrm{U}(f)-M_\mathrm{g}^\mathrm{ref}(f^\mathrm{ref}),
  \label{eq:ext}
\end{equation}
\ie the energy released when the remnant spins down from $f$ to $f^\mathrm{ref}$ at fixed baryon mass.
It is the second of the two reservoirs used in Sec.~\ref{sec:injection}, namely the energy that the remnant can still give away once the differential rotation is lost, whereas $E^\mathrm{D}_\mathrm{brake}$ of Eq.~\eqref{eq:free-energy-final} is the energy released while the differential rotation is being lost.

As the rotation slows down, the kinetic energy decreases; in addition, the star generally contracts, so that the gravitational binding energy is released and adds to what the kinetic energy alone provides.
Since the gravitational mass includes both contributions, using the kinetic energy difference alone to estimate the extractable energy~\cite{Metzger2015, Beniamini2021, Tang2025a} in the braking process will underestimate it.
The underestimate increases with $M_\mathrm{b}$ (see Fig.~\ref{fig:mgspr}): it reaches up to $\sim 20\%$ at $M_\mathrm{b}=M_\mathrm{b}^\mathrm{TOV}$ and even exceeds $100\%$ at the high-mass end of $M_\mathrm{b}>M_\mathrm{b}^\mathrm{TOV}$ (see a detailed discussion in Appendix~\ref{app:extractable}).
These differences highlight the importance of using the gravitational mass difference to calculate the extractable energy rather than the kinetic energy in rotating NSs with $M_{\mathrm{b}}>M_{\mathrm{b}}^{\mathrm{TOV}}$, where most BNS merger remnants sit.

\subsection{Angular momentum and mass loss}
\label{sec:loss}

Four main processes transport angular momentum in a post-merger remnant: GW emission, mass outflow, neutrino emission, and magnetic stresses~\cite{Ma2017, Shibata2019a, Shao2020, Fan2023, ChenY2026}.
Only the first three directly carry energy and angular momentum \emph{away} from the remnant, and therefore only these processes enter the balance equations of Sec.~\ref{sec:budget}.
The magnetic stresses, by contrast, do not themselves carry energy or angular momentum away from the remnant; instead, they act inside it and assist the outward transport by facilitating some of the mass and angular momentum loss processes above.
Their internal redistribution of angular momentum is what relaxes the remnant toward uniform rotation and converts the differential braking energy $E^\mathrm{D}_\mathrm{brake}$ of Eq.~\eqref{eq:free-energy-final} into magnetic energy.

We therefore track a merger remnant driven by the three outward angular momentum loss channels (internal transport is treated in Ref.~\cite{Margalit2022}, and both are treated in Ref.~\cite{Ng2026b}). The track starts at the reference time $t_\mathrm{s}$, at which the remnant is taken from the simulation, and ends on a configuration with the same baryon mass and angular momentum constructed from cold EOS tables by \rns. The differential braking energy of Eq.~\eqref{eq:free-energy-final} at that endpoint is the budget left for the amplification of the magnetic field. In what follows, we describe the angular momentum loss ratio of each channel, the upper limit of the energy it can carry away, and the construction of possible evolution tracks of the remnant.

\subsubsection{Loss ratios}

Each mass and angular momentum loss channel is fully described by the two loss ratios it imposes on the evolution, \ie the slopes of the track in the $(J,M_\mathrm{g},M_\mathrm{b})$ space: the energy it takes away per unit angular momentum, $\mathrm{d}M_\mathrm{g}/\mathrm{d}J$, and the baryon mass it takes with it, $\mathrm{d}M_\mathrm{b}/\mathrm{d}J$.
Following Ref.~\cite{Shibata2019a}, we adopt the following, slightly modified, loss ratios:
\begin{equation}
  \label{eq:Jloss}
  \begin{aligned}
  \mathrm{GW:}\quad
  &\frac{\mathrm{d}M_\mathrm{g}}{\mathrm{d}J}=\pi f_2,
  &&\frac{\mathrm{d}M_\mathrm{b}}{\mathrm{d}J}=0,\\[2pt]
  \nu:\quad
  &\frac{\mathrm{d}M_\mathrm{g}}{\mathrm{d}J}=\frac{3}{2R^2\Omega F^{\nu}_\mathrm{c}},
  &&\frac{\mathrm{d}M_\mathrm{b}}{\mathrm{d}J}=0,\\[2pt]
  \mathrm{ejecta:}\quad
  &\frac{\mathrm{d}M_\mathrm{g}}{\mathrm{d}J}=\frac{\Gamma_\mathrm{eje}}{\sqrt{M_\mathrm{g}R_\mathrm{torus}}},
  &&\frac{\mathrm{d}M_\mathrm{b}}{\mathrm{d}J}=\frac{1}{\sqrt{M_\mathrm{g}R_\mathrm{torus}}}.
  \end{aligned}
\end{equation}
We have assumed that the two radiative channels carry energy and angular momentum in a fixed ratio,
\begin{equation}
  \label{eq:Jnu}
  J_\mathrm{GW}=\frac{E_\mathrm{GW}}{\pi f_2},
  \qquad
  J_\nu= F^{\nu}_\mathrm{c}\,(\tfrac{2}{3}R^2\Omega)\,E_\nu .
\end{equation}
They leave the baryon mass untouched, so only $M_\mathrm{g}$ and $J$ are reduced in the two radiative channels.
The symbols are as follows.
$f_2$ is the dominant frequency of the post-merger GW spectrum.
We take it from Table~III of Ref.~\cite{Han2025}, and compute it using the same algorithm for the additional three cases with $\Gamma_\mathrm{th}=1.75$ that are not listed there.
Here $R$ and $\Omega$ are evaluated on the equator of the isodensity surface $\rho=\rho_\mathrm{cut}$ at $t_\mathrm{s}$ and averaged over azimuth.
In addition, we assume
\begin{equation}
  \label{eq:Jdef}
  R_\mathrm{torus}\equiv\frac{J_\mathrm{torus}^2}{M_\mathrm{b,torus}^2}\frac{1}{M_\mathrm{g}(t_\mathrm{s})},
  \qquad
  \frac{\mathrm{d}M_\mathrm{g}}{\mathrm{d}M_\mathrm{b}}=\Gamma_\mathrm{eje},
\end{equation}
with $M_\mathrm{b,torus}$ and $J_\mathrm{torus}$ (see Appendix~\ref{append1:torus}) the baryon mass and the angular momentum of the bound matter between $10^7\,\mathrm{g\,cm^{-3}}$ and $\rho_\mathrm{cut}$ of the simulation~\cite{Fujibayashi2017}.
We take $\Gamma_\mathrm{eje}$ equal to the Lorentz factor of the Keplerian motion at $R_\mathrm{torus}$ at the selected time, $\Gamma_\mathrm{eje}=1/\sqrt{1-M_\mathrm{g}(t_\mathrm{s})/R_\mathrm{torus}}$.
$R_\mathrm{torus}$ is the radius of the Newtonian circular orbit that carries the specific angular momentum of the torus, so it is the areal radius that the Keplerian relation above requires, and being built from $(J_\mathrm{torus}/M_\mathrm{b,torus})^2/M_\mathrm{g}$ it does not depend on the choice of spatial coordinates.
Finally, the right part of Eq.~\eqref{eq:Jdef} holds only if the lost mass is small compared with $M_\mathrm{g}$ and the outflow is in the far-field, low-velocity region, both of which hold here.

Special attention should be paid to the coefficient $F^{\nu}_\mathrm{c}$ in Eq.~\eqref{eq:Jloss}, which is not directly taken over from previous works.
The neutrino contribution follows from Eq.~(8) of Ref.~\cite{Baumgarte1998}, $\dot J_{\nu}=-\langle\varpi^2 \rangle \Omega L_{\nu}$, where $L_\nu$ is the neutrino luminosity and the average is over the surface from which the radiation escapes.
Only for a uniformly rotating sphere do we have $\langle\varpi^2 \rangle \Omega = \tfrac23 R^2 \Omega$, and this value decreases slightly for the uniformly rotating ellipsoid~\cite{Baumgarte1998}.
We retain the form of this relation, generalize it to differential rotation, and measure the coefficient that it leaves free (fifth column of Table~\ref{tab:transport}),
\begin{equation}
  \label{eq:Fnu}
  \langle g_{\varphi\varphi}\Omega \rangle=\frac{2}{3}R^2\Omega\,F^{\nu}_\mathrm{c},
  \qquad
  F^{\nu}_\mathrm{c}=\frac{\oint g_{\varphi\varphi}\Omega(\theta, \varphi)\,\mathrm{d}A}{\tfrac23R^2\Omega \oint \mathrm{d}A},
\end{equation}
with $\mathrm{d}A$ the proper area element of the isodensity surface $\rho=\rho_\mathrm{cut}$.
From Table~\ref{tab:transport} we observe that using the uniformly-rotating-sphere value $\tfrac23R^2\Omega$ of Refs.~\cite{Baumgarte1998, Shibata2019a} underestimates the angular momentum carried away by neutrinos by about $20\%$--$30\%$.
We note that this coefficient would differ for a different cut-off density, and that the true surface would be where the neutrino optical depth is $\tau_{\nu}\sim 1$, which is not necessarily the density cut chosen here; however, given the current lack of accurate knowledge of the neutrino radiation process in the merger remnant and the purely hydrodynamic nature of simulations used in this work, we adopt a simplified prescription here.

\subsubsection{Upper limits of loss}
We now turn to the energy budget of each mass and angular momentum loss channel, as no channel can accommodate an arbitrary amount of loss.
Most of the physical considerations below are motivated by the simulation results.
Note that, although these values are reasonable, they are nevertheless assumptions, and that each budget is the loss still available after $t_\mathrm{s}$, so that the energy and angular momentum already carried away during the inspiral, the merger, and the earliest post-merger stage before $t_\mathrm{s}$ are not part of it.

\begin{itemize}[leftmargin=*]
\item \emph{GWs} carry away no more energy than was radiated between the merger and the selected time, \ie
\begin{equation}
  \label{eq:Egwmax}
  E_\mathrm{GW}^{\max}= \int_{t_\mathrm{merge}}^{t_\mathrm{s}} \frac{\mathrm{d}E_\mathrm{GW}}{\mathrm{d}t} \, \mathrm{d}t,
\end{equation}
where $E_\mathrm{GW}$ is the radiated GW energy measured at the coordinate radius $r_\mathrm{ext}=300\Msun$. In this way, the GW amplitude does not need to be extrapolated, and no assumption has to be made about what damps the GW. While other estimates of the post-merger GW output exist, they rest on rather short simulations~\cite{Bernuzzi2015, Zappa2017, ChenY2026} and are consistent with the ceiling adopted here. Over the long term, further angular momentum may be lost through the GW channel due to the $r$-mode instability, the low-$T/W$ instability, and so on, which our simulations cannot capture.

\item \emph{Neutrinos} radiate no more than the thermal energy the remnant still holds, \ie
\begin{equation}
  \label{eq:Enumax}
  E_\nu^{\max} = E_\mathrm{th},
\end{equation}
where $E_\mathrm{th}$ is defined in Eq.~\eqref{eq:app_Eth}.
This is the simplest of the three budgets, as it is the thermal energy of the remnant itself, taken from the simulation at the same density cut that defines $M_\mathrm{b}$, $J$, and $M_\mathrm{g}$.

\item \emph{Ejecta} carry away no more than $0.1\Msun$, \ie
\begin{equation}
  \label{eq:Mejemax}
  M_\mathrm{eje}^{\max} = 0.1\Msun.
\end{equation}
Neutrino irradiation and the magnetic (or viscous) stresses dominate the post-merger ejection.
Both redistribute angular momentum and can change the density profile fast enough to launch shocks, and each drives an outflow of its own.
Their relative weight is still under debate, as neither has the amplification of the magnetic field and its instabilities been fully captured by simulations, nor have 3D, long-term, high-resolution, multi-species neutrino transport simulations been performed.
Although both play a role in the mass ejection from the remnant and the torus, and it is very hard to single out the effect of each, here we use a crude approximation to separate them, in which magnetic effects dominate the further mass ejection from eruptions and magnetic-threading outflows from the remnant, whereas neutrinos dominate the torus mass ejection through irradiation and cooling.
From the magnetic side, various works have shown that magnetic fields can drive strong outflows~\cite{Kiuchi2011, Siegel2014, Most2023a, Musolino2024b, Jiang2025, Fields2025, Wen2026}.
Here we take $0.1\Msun$~\cite{Ciolfi2019, Kiuchi2023, Jiang2025}, which is also consistent with an effective shear viscosity $\alpha_\mathrm{vis}\approx0.2$~\cite{Bamber2024b} combined with the empirical relation $M_\mathrm{eje}\sim0.01\Msun(\alpha_\mathrm{vis}/0.02)$ from Refs.~\cite{Fujibayashi2017, Shibata2019b}.
From the neutrino side, a torus of $0.25\Msun$ can be formed~\cite{Martin2015, Radice2018b, Vincent2019, Fujibayashi2020, Radice2023}, of which $20\%$~\cite{Radice2018b, Vincent2019}, \ie $0.05\Msun$, could be unbound in the end purely as an effect of neutrinos (a larger fraction of $30\%$--$50\%$ is expected in Refs.~\cite{Fujibayashi2017, Shibata2019b, Kawaguchi2020, Fujibayashi2020}).
Since we set a cut-off density, beyond which the mass and angular momentum of a torus are excluded from the remnant, our budget for further mass ejection should not include the torus part.
Thus we only have an upper limit of $0.1\Msun$ that will be ejected further from the central remnant; the part of the torus that is not ejected will finally be accreted onto the remnant during the long-term cooling, but its timescale is much longer than the timescale of magnetic field amplification by differential rotation, so we exclude it throughout this work.
Nevertheless, even excluding the contribution of mass ejection from the torus, an upper limit of $0.1\Msun$ already equals what long-term 3D high-resolution simulations including both realistic neutrino transport and magnetic shear give~\cite{Kiuchi2023, Kiuchi2026b}; we therefore regard the upper limit of further mass loss from the remnant adopted here as realistic.
Finally, we caution that this value could also change if additional particle species such as muons and pions played a role in the merger~\cite{Ng2024, Pajkos2024, Gieg2026b}.
\end{itemize}

\subsubsection{Evolution tracks and statistics}

We construct a set of tracks and use their statistics to accommodate our ignorance of the true evolutionary trajectory that a realistic merger remnant will follow.
An evolution track is defined by the weight $w_i\ge0$ of the angular momentum loss ratio that each channel takes; every weight is taken from $\{1, 1/2, 1/3\}$, the weights of a track add up to one, and they are changed where a channel cannot pay its share (see below).
The track is the integral of the combined slopes
\begin{equation}
  \label{eq:Jtrack}
  \frac{\mathrm{d}M_\mathrm{b}}{\mathrm{d}J}=\sum_iw_i\left(\frac{\mathrm{d}M_\mathrm{b}}{\mathrm{d}J}\right)_i,
  \,
  \frac{\mathrm{d}M_\mathrm{g}}{\mathrm{d}J}=\sum_iw_i\left(\frac{\mathrm{d}M_\mathrm{g}}{\mathrm{d}J}\right)_i
\end{equation}
from the remnant of the simulation downward to uniformly rotating configurations with variable $J$.
Along a track the loss ratios of Eq.~\eqref{eq:Jloss} are held fixed, and each family of loss histories is integrated explicitly.

\begin{table*}[!t]
\centering
\caption{Properties of the merger remnants at $t_\mathrm{s}$ and of the endpoints of their evolution tracks (one row per run). A run is named by its EOS (L, M, H), by the initial binary total mass in tenths of a solar mass (M27 for $2.7\Msun$), and by its thermal index (G1.5 for $\Gamma_\mathrm{th}=1.5$). Columns: the selected time $t_\mathrm{s}$ of the first 3D snapshot after $t_\mathrm{cross}$ (Eq.~\eqref{eq:ts}); the energy ceilings of the two radiative channels (Sec.~\ref{sec:loss}); the coefficient $F^{\nu}_\mathrm{c}$ of Eq.~\eqref{eq:Fnu}; the number of the seven evolutions that reach the turning-point line and are taken to collapse there; the differential braking energy $E^\mathrm{D}_\mathrm{brake}$ of Eq.~\eqref{eq:free-energy-final}; the rotation frequency $f_\mathrm{rem}$ of the uniformly rotating configuration; the ratio $T_\mathrm{kin}/E_\mathrm{bind}$ of that rotation to the binding energy of the same configuration, in units of $0.1$; the uniform braking energy $E^\mathrm{U}_\mathrm{brake}$ of that configuration, Eq.~\eqref{eq:ext}; and, to the right of the rule, what follows if the whole differential braking energy is transferred into the magnetic field, $E^\mathrm{D}_\mathrm{brake}=E_\mathrm{mag}$ (all the following quantities are discussed in Sec.~\ref{sec:injection}): the field $\bar{B}_{\max}$ (proper volume average, see Eq.~\eqref{eq:Bmax}) it would generate, the largest electric field of that rotating dipole in units of the critical field of quantum electrodynamics (QED), Eq.~\eqref{eq:Emax}, the dipole timescale (see Eq.~\eqref{eq:tdip}) over which that field spends $E^\mathrm{U}_\mathrm{brake}$ with $\eta_\mathrm{dip}=1$ in Eq.~\eqref{eq:Bp}, and the Hall drift timescale of the crust at that strength of magnetic field, Eq.~\eqref{eq:thall}, for a drift length of $L=1\,\mathrm{km}$. Entries are means over the evolutions that count, with their standard deviations, except the electric field ratio and the two timescales, which are quoted without a spread. Finally, a dash marks a run in which all tracks collapse.}
\label{tab:transport}
\footnotesize
\setlength{\tabcolsep}{2.5pt}%
\renewcommand{\arraystretch}{1.08}%
\resizebox{\ifdim\width>\linewidth\linewidth\else\width\fi}{!}{%
\begin{tabular}{@{}l c c cc c c c cc| c c c c@{}}
\hline\hline
Model & $t_\mathrm{s}-t_\mathrm{merge}$ & $E_\mathrm{GW}^{\max}$ & $E_\nu^{\max}$ & $F^{\nu}_\mathrm{c}$ & $N_\mathrm{collapse}$ & $E^\mathrm{D}_\mathrm{brake}$ & $f_\mathrm{rem}$ & $T_\mathrm{kin}/E_\mathrm{bind}$ & $E^\mathrm{U}_\mathrm{brake}$ & $\bar{B}_{\max}$ & $\frac{E^\mathrm{el}_{\max}}{E^\mathrm{el}_\mathrm{cr}}$ & $t_\mathrm{dip}$ & $t_\mathrm{Hall}$ \\
 & $[\mathrm{ms}]$ & $[10^{-2}M_\odot]$ & $[10^{-2}M_\odot]$ &  & $[\mathrm{of}\ 7]$ & $[10^{52}~\mathrm{erg}]$ & $[\mathrm{kHz}]$ & $[0.1]$ & $[10^{52}~\mathrm{erg}]$ & $[10^{17}~\mathrm{G}]$ &  & $[\mathrm{ms}]$ & $[\mathrm{yr}]$ \\
\hline
L-M27-G2.0 & 6.3 & 5.87 & 6.15 & 1.25 & 7 & -- & -- & -- & -- & -- & -- & -- \\
M-M27-G1.1 & 15.8 & 3.50 & 7.22 & 1.31 & 0 & $6.96\pm 0.54$ & $1.37\pm 0.05$ & $1.19\pm 0.16$ & $6.50\pm 2.25$ & $3.82\pm 0.07$ & $1.0\times10^{3}$ & 1.29 & $1.7\times10^{2}$ \\
M-M27-G1.2 & 11.1 & 3.72 & 6.33 & 1.31 & 0 & $5.78\pm 0.86$ & $1.31\pm 0.06$ & $1.05\pm 0.16$ & $5.93\pm 2.53$ & $3.57\pm 0.23$ & $7.7\times10^{2}$ & 2.15 & $1.8\times10^{2}$ \\
M-M27-G1.3 & 14.8 & 3.27 & 6.00 & 1.32 & 0 & $4.93\pm 0.96$ & $1.30\pm 0.06$ & $1.05\pm 0.16$ & $7.61\pm 2.42$ & $3.28\pm 0.29$ & $6.9\times10^{2}$ & 3.61 & $2.0\times10^{2}$ \\
M-M27-G1.4 & 13.3 & 3.86 & 5.46 & 1.29 & 0 & $4.86\pm 1.04$ & $1.25\pm 0.08$ & $0.91\pm 0.17$ & $5.74\pm 2.67$ & $3.33\pm 0.33$ & $5.8\times10^{2}$ & 3.90 & $1.9\times10^{2}$ \\
M-M27-G1.5 & 12.8 & 3.77 & 5.18 & 1.29 & 0 & $4.88\pm 1.18$ & $1.28\pm 0.07$ & $0.97\pm 0.17$ & $6.20\pm 2.78$ & $3.30\pm 0.38$ & $6.3\times10^{2}$ & 3.61 & $2.0\times10^{2}$ \\
M-M27-G1.6 & 16.0 & 4.38 & 4.68 & 1.33 & 0 & $2.92\pm 2.27$ & $1.26\pm 0.09$ & $0.87\pm 0.22$ & $4.17\pm 3.36$ & $2.28\pm 1.31$ & $3.8\times10^{2}$ & 4.89 & $7.5\times10^{2}$ \\
M-M27-G1.7 & 15.8 & 4.12 & 4.50 & 1.27 & 0 & $3.55\pm 1.94$ & $1.27\pm 0.09$ & $0.96\pm 0.22$ & $5.80\pm 2.76$ & $2.72\pm 0.93$ & $4.9\times10^{2}$ & 6.56 & $2.8\times10^{2}$ \\
M-M27-G1.75 & 15.0 & 3.90 & 4.13 & 1.28 & 0 & $3.68\pm 2.04$ & $1.32\pm 0.05$ & $1.04\pm 0.15$ & $5.57\pm 3.07$ & $2.71\pm 0.89$ & $5.9\times10^{2}$ & 4.36 & $2.8\times10^{2}$ \\
M-M27-G1.8 & 17.8 & 4.29 & 4.14 & 1.31 & 0 & $3.35\pm 2.07$ & $1.29\pm 0.09$ & $0.92\pm 0.22$ & $4.62\pm 3.45$ & $2.66\pm 0.90$ & $4.8\times10^{2}$ & 3.90 & $2.7\times10^{2}$ \\
M-M27-G1.9 & 15.0 & 3.65 & 4.04 & 1.23 & 0 & $4.14\pm 1.95$ & $1.31\pm 0.05$ & $1.02\pm 0.16$ & $5.63\pm 3.06$ & $2.95\pm 0.74$ & $6.2\times10^{2}$ & 3.38 & $2.3\times10^{2}$ \\
M-M27-G2.0 & 13.5 & 4.48 & 4.02 & 1.20 & 0 & $2.89\pm 2.15$ & $1.28\pm 0.08$ & $0.93\pm 0.23$ & $4.53\pm 3.64$ & $2.40\pm 1.04$ & $4.3\times10^{2}$ & 4.93 & $3.3\times10^{2}$ \\
M-M28-G1.75 & 20.7 & 5.56 & 5.60 & 1.25 & 4 & $3.08\pm 2.72$ & $1.44\pm 0.04$ & $1.02\pm 0.09$ & $1.02\pm 1.56$ & $2.37\pm 1.41$ & $5.3\times10^{2}$ & 2.57 & $4.5\times10^{2}$ \\
M-M29-G1.5 & 13.2 & 4.60 & 7.37 & 1.27 & 6 & $4.86\pm 2.49$ & $1.59\pm 0.05$ & $1.25\pm 0.04$ & 0.28 & $3.22\pm 0.97$ & $1.1\times10^{3}$ & 0.06 & $2.2\times10^{2}$ \\
M-M29-G1.6 & 10.4 & 4.73 & 6.64 & 1.28 & 6 & $4.30\pm 2.36$ & $1.56\pm 0.06$ & $1.19\pm 0.03$ & 0.28 & $3.04\pm 1.02$ & $9.1\times10^{2}$ & 0.08 & $2.5\times10^{2}$ \\
M-M29-G1.7 & 24.4 & 5.69 & 6.92 & 1.24 & 6 & $6.88\pm 0.93$ & $1.59\pm 0.03$ & $1.21\pm 0.07$ & $2.1\times10^{-2}$ & $4.06\pm 0.25$ & $1.3\times10^{3}$ & $6.2\times10^{-3}$ & $1.6\times10^{2}$ \\
M-M29-G1.75 & 22.8 & 5.15 & 6.40 & 1.23 & 6 & $5.13\pm 1.90$ & $1.58\pm 0.05$ & $1.21\pm 0.04$ & 0.24 & $3.43\pm 0.68$ & $1.1\times10^{3}$ & 0.07 & $1.9\times10^{2}$ \\
M-M29-G1.8 & 14.5 & 5.71 & 5.96 & 1.22 & 6 & $6.14\pm 1.32$ & $1.59\pm 0.05$ & $1.23\pm 0.05$ & 0.12 & $3.79\pm 0.42$ & $1.2\times10^{3}$ & 0.04 & $1.7\times10^{2}$ \\
M-M29-G1.9 & 17.3 & 5.92 & 5.97 & 1.18 & 6 & $7.22\pm 0.99$ & $1.61\pm 0.03$ & $1.24\pm 0.07$ & $3.1\times10^{-3}$ & $4.13\pm 0.25$ & $1.4\times10^{3}$ & $8.7\times10^{-4}$ & $1.5\times10^{2}$ \\
M-M29-G2.0 & 13.4 & 5.53 & 5.51 & 1.17 & 6 & $5.93\pm 1.78$ & $1.61\pm 0.06$ & $1.27\pm 0.04$ & 0.19 & $3.68\pm 0.57$ & $1.3\times10^{3}$ & 0.05 & $1.8\times10^{2}$ \\
H-M27-G1.1 & 20.6 & 2.40 & 5.01 & 1.34 & 0 & $5.49\pm 0.62$ & $1.09\pm 0.04$ & $1.21\pm 0.14$ & $12.26\pm 1.32$ & $2.80\pm 0.09$ & $6.2\times10^{2}$ & 4.94 & $2.3\times10^{2}$ \\
H-M27-G1.5 & 28.1 & 3.08 & 3.74 & 1.30 & 0 & $2.89\pm 1.65$ & $0.99\pm 0.08$ & $0.93\pm 0.21$ & $9.78\pm 2.21$ & $2.01\pm 0.80$ & $2.7\times10^{2}$ & 34.40 & $4.4\times10^{2}$ \\
H-M27-G2.0 & 15.3 & 2.36 & 3.08 & 1.28 & 0 & $3.36\pm 1.46$ & $1.07\pm 0.04$ & $1.13\pm 0.13$ & $11.95\pm 1.60$ & $2.16\pm 0.52$ & $4.2\times10^{2}$ & 12.37 & $3.2\times10^{2}$ \\
H-M29-G1.1 & 20.0 & 3.28 & 6.73 & 1.32 & 1 & $6.98\pm 0.36$ & $1.15\pm 0.04$ & $1.16\pm 0.16$ & $5.03\pm 2.33$ & $3.26\pm 0.10$ & $7.3\times10^{2}$ & 1.35 & $2.0\times10^{2}$ \\
H-M29-G1.5 & 18.6 & 3.89 & 4.91 & 1.29 & 1 & $4.64\pm 1.01$ & $1.04\pm 0.07$ & $0.83\pm 0.18$ & $3.16\pm 2.58$ & $2.79\pm 0.28$ & $4.0\times10^{2}$ & 2.81 & $2.3\times10^{2}$ \\
H-M29-G2.0 & 16.7 & 3.47 & 3.66 & 1.26 & 0 & $3.84\pm 1.83$ & $1.13\pm 0.05$ & $1.07\pm 0.19$ & $4.78\pm 2.81$ & $2.39\pm 0.61$ & $4.7\times10^{2}$ & 3.17 & $2.9\times10^{2}$ \\
\hline\hline
\end{tabular}}
\end{table*}

\begin{figure*}[!t]
  \includegraphics[width=0.98\linewidth]{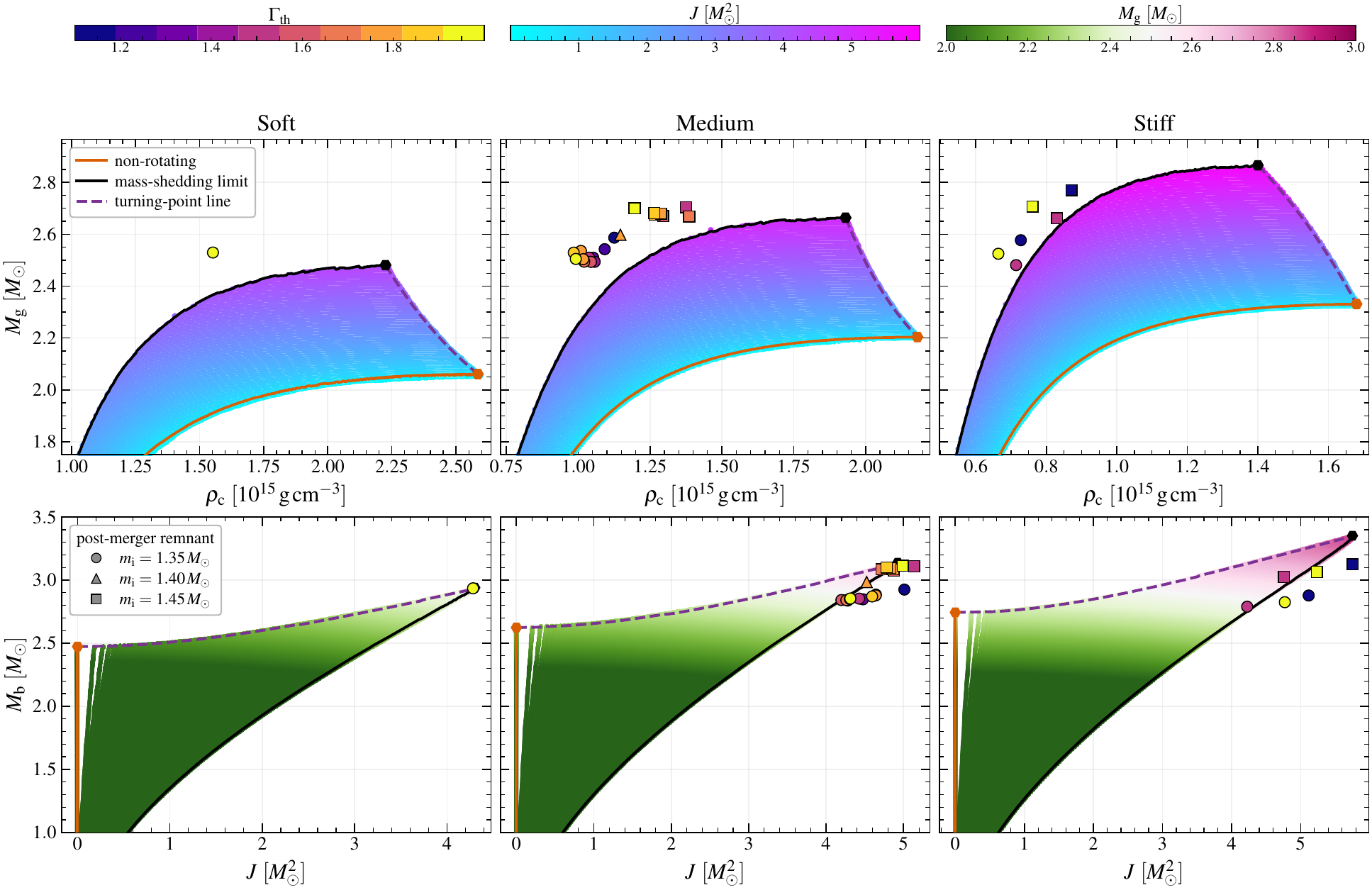}
  \caption{Equilibrium configurations of the three cold EOSs (soft, medium
  and stiff, from left to right) in the gravitational mass--central density plane (top row) and in
  the baryon mass--angular momentum plane (bottom row). The non-rotating
  sequence (solid orange), the mass-shedding limit (solid black) and the
  turning-point line (dashed violet) keep the same style in every figure of this
  paper.
  In the top row the thin curves are sequences of constant angular momentum,
  colored by $J$ (scale at the top middle), each drawn up to the turning-point line. In the bottom row the cold series is colored by its gravitational
  mass (scale at the top right). The turning-point line connecting the TOV mass and
  maximum mass configurations (hexagons) is also mapped onto this plane. The
  symbols are the post-merger remnants of this work that appear in
  Table~\ref{tab:transport}, at their own central rest-mass density and
  with $M_\mathrm{b}$, $M_\mathrm{g}$, and $J$ taken as the volume integrals inside
  $\rho_\mathrm{cut}=10^{11}~\mathrm{g\,cm^{-3}}$ for the selected time $t_\mathrm{s}$, with the
  shape of them giving the initial binary mass and the color the $\Gamma_\mathrm{th}$ of the run
  (scale at the top left).}
  \label{fig:stablemap}
\end{figure*}

The three angular momentum loss channels can be combined in seven ways, each representing a different split of the angular momentum loss among the channels:
(1) the pure tracks, which give the whole share to a single channel, with GWs (i-G), neutrinos (i-N), or the ejecta (i-M) alone;
(2) the half tracks, in which two channels act on comparable, much shorter timescales and share the loss equally, with waves and neutrinos (ii-GN), waves and ejecta (ii-GM), or neutrinos and ejecta (ii-NM), each of the two taking the weight $1/2$;
and (3) the one-third track, in which the timescales of the three are similar and each takes a weight of $1/3$ (iii-GNM).
These are the weights a track starts with, where a channel that has spent its allowance drops out of the split, and its share goes equally to the channels that still have some, so that the weights still add up to one.
These different combinations can be understood in two ways: the first is in terms of timescale, and the second in terms of significance.
The channel considered in each track can be seen as acting faster than the others, or as the others carrying much less angular momentum than the upper limits assumed here in a realistic evolution.
We note that holding the loss ratios and weights fixed until each channel has used up its allowance is an idealization; however, with no prior knowledge of how the relative strength and timescale of the three channels evolve in such complex dynamics, we can only adopt a simplified prescription and combination.

We follow every evolution track until it uses up all the angular momentum loss budget of its channels or reaches the turning-point line, beyond which (at lower $J$) the configuration is secularly unstable.
Although collapse is not guaranteed there, the onset of dynamical instability is close, as its central density differs from that of the turning point by less than $1\%$ in most cases~\cite{Gao2026_res}.
We therefore record a track that reaches the line as a collapse.
An evolution that never reaches the uniformly rotating region does not count, since there is no uniform configuration to compare with.
At the endpoint, the differential braking energy $E^\mathrm{D}_\mathrm{brake}$ of Eq.~\eqref{eq:free-energy-final} is what is left for the magnetic field amplification process.
The statistics (the mean and standard deviation) are computed from the endpoints of every track and are collected in Table~\ref{tab:transport}.

Throughout this work, the cold uniformly rotating configuration is used as a reference state, rather than as the thermodynamic endpoint of every loss history.
In the absence of efficient cooling cases, part of the energy excess above this reference remains in thermal form, \ie contributing to $E_\mathrm{res}$.
Our estimates of the magnetic field amplification should therefore be interpreted as the most optimistic energetic bounds in the amplification processes, not as predictions of the exact magnetic energy attained.
However, we also note that a possible convective dynamo might contribute to amplifying the magnetic field by converting part of the thermal energy (in $E_{\mathrm{res}}$) into convective kinetic energy and then into magnetic energy~\cite[see, \eg][]{Duncan1992, Rheinhardt2005, Naso2007, Verma2026}.
This further validates the upper-limit approach adopted here, except that this transfer efficiency may be low or even absent in some cases.
Since detailed neutrino dynamics is needed to fully investigate this, we leave the problem of the efficiency of transferring residual thermal energy into magnetic energy to future work.

\section{Results}
\label{sec:results}

We describe the results starting from the remnant properties, then the tracks followed during the angular momentum loss, and finally the endpoint of each track, which is compared with the uniformly rotating configuration; the impact of the thermal index, total mass, and EOS is then investigated. Last, we discuss the implications for GRB physics and for the long-term stability and dissipation of the magnetic field.

\subsection{Remnant properties at the selected time}
\label{sec:remnant}

In this subsection we focus on the remnant properties listed in Table~\ref{tab:transport} before any angular momentum is lost (the loss process is considered in the next subsection).
The second column of Table~\ref{tab:transport} is $t_\mathrm{s}$, \ie the time of the 3D snapshot selected immediately after $t_\mathrm{cross}$ (see Eq.~\eqref{eq:ts}).
It depends more strongly on the EOS than on the binary mass, \ie the stiff runs have longer $t_\mathrm{s}$, as a larger remnant readjusts more slowly (also see the right panel of Fig.~\ref{fig:eos}).
The thermal index, by contrast, shows no systematic trend.

The third and fourth columns of Table~\ref{tab:transport} present the two ceilings for the radiative channels.
The GW ceiling shows no systematic trend with the thermal index, whereas the neutrino ceiling decreases with increasing $\Gamma_{\mathrm{th}}$, because an excess of thermal pressure expels matter further out to less dense regions, leaving less total thermal energy for the remnant.
On the other hand, the two ceilings share the same trend in two respects.
First, they increase with total mass, as a larger content of matter has stronger GW emission and sustains more thermal energy.
Second, they both increase if the EOS becomes softer, as a softer EOS produces more compact remnants that emit more GWs and oscillate more violently, transferring more kinetic energy into thermal energy.

We then turn to the cold equilibrium configurations that uniform rotation can reach and to the position of the merger remnants among them; both are shown in Fig.~\ref{fig:stablemap}.

In the top row, the reachable set is the region between the non-rotating sequence (orange) and the mass-shedding sequence (black), the fastest uniform configuration that a given baryon mass allows.
Sequences of constant angular momentum foliate it, with each thin curve ending on the turning-point line (dashed violet), beyond which no long-term stable configuration with that $J$ exists.
Regarding the EOS effect, we observe that rotation increases the maximum supported mass the most in the stiff case (see also Table~\ref{tab:EOSs}), and that the range of central densities of the stiff EOS is both lower and narrower than that of the softer ones.

The bottom row of Fig.~\ref{fig:stablemap} shows the same configurations again in the baryon mass--angular momentum plane, the plane in which the evolutions of Sec.~\ref{sec:transport} are drawn.
The set is bounded by the same two sequences, which here read as the two limits of rotation at a fixed baryon mass, the mass-shedding sequence being the largest $J$ that a uniform rotator of that $M_\mathrm{b}$ can hold and the turning-point line the smallest, so that the horizontal width of the set at a given $M_\mathrm{b}$ is the range of rotation that uniform rotation can reach.
Being colored by gravitational mass, the cold series in this row shows that a stiff EOS can sustain a higher baryon mass and angular momentum, with the maximum angular momentum $\sim 5.7 \Msun^2$ for the stiff EOS and $\sim 4.2 \Msun^2$ for the soft one.

\begin{figure*}[!t]
  \includegraphics[width=0.98\linewidth, height=0.65\linewidth]{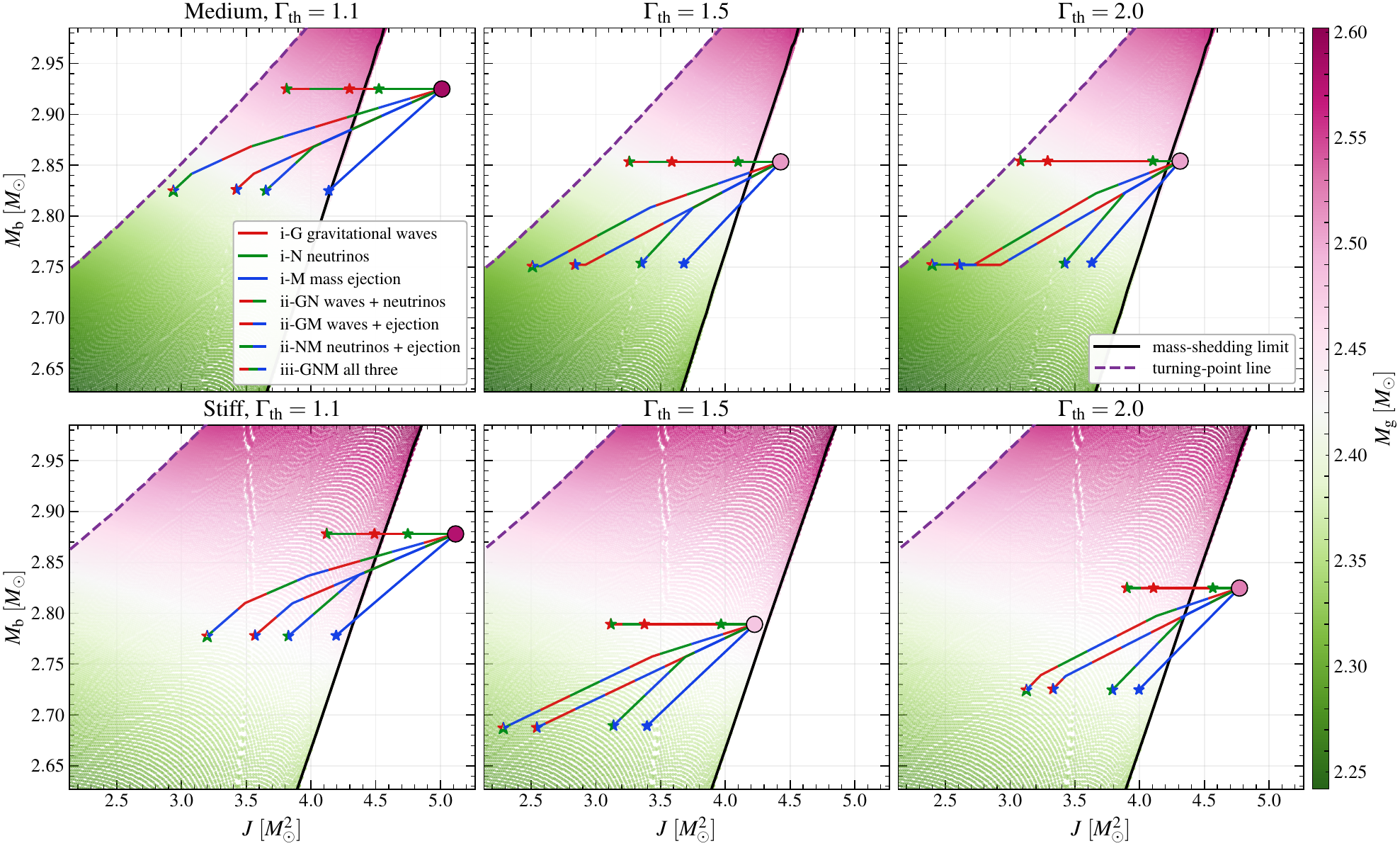}
  \caption{The seven evolution tracks of the lightest binary ($M_\mathrm{tot}=2.7\Msun$) in a grid of EOS and thermal index, the medium EOS in the top row and the stiff one in the bottom row, with $\Gamma_\mathrm{th}=1.1$, $1.5$, and $2.0$ from left to right, one run per panel.
  Each panel shows the $J$--$M_\mathrm{b}$ plane of its EOS, with the equilibrium configurations colored by gravitational mass (color scale on the right).
  Both axes and the color scale are shared by the six panels, allowing the tracks to be compared from one panel to the next.
  The legend of the upper-left panel indicates how the angular momentum loss is shared: the three pure tracks i-G, i-N, and i-M; the three pairs ii-GN, ii-GM, and ii-NM; and iii-GNM with all three channels combined.
  The large circle denotes the remnant at the selected time, with its face color indicating the gravitational mass at that moment.
  A filled star marks the end of a track whose allowances are exhausted, and the marker of a track that mixes channels is divided into the colors of those channels, as is the line leading to it.}
  \label{fig:transport}
\end{figure*}

The merger remnants sit as points in Fig.~\ref{fig:stablemap}.
The central density of a merger remnant is mostly smaller than that of the uniform configuration with the same gravitational mass, owing to the centrifugal effect of the differential rotation and the excess of thermal pressure.
We also observe that higher-mass remnants have a larger central density and a higher angular momentum.
Most of the remnants are super-Keplerian with respect to the equilibrium sequences~\cite{Bernuzzi2015, Radice2018a}, thus further angular momentum loss is needed to reach uniform rotation.
There are, however, also seven of these remnants that already sit inside the region allowed by uniform rotation.
A special case among these seven is the soft EOS one, since the chosen mass is a bit large for the soft EOS and most of the models collapse within $25~\mathrm{ms}$ after the merger~\cite{Han2025} and are not included in Table~\ref{tab:transport}, except in the largest $\Gamma_\mathrm{th}$ case, where it sits exactly on the mass-shedding limit of the maximum-mass configuration.
We note that this configuration will eventually collapse after further losing angular momentum from any possible channel, which is also the reason that the soft EOS is not shown in Fig.~\ref{fig:transport} and later figures.

As for the thermal effect, a remnant with a larger $\Gamma_\mathrm{th}$ has a lower central density, owing to the excess of thermal pressure support.
The remnant with the smallest $\Gamma_\mathrm{th}$ has the highest gravitational (and baryon) mass and angular momentum, while an intermediate $\Gamma_\mathrm{th}$ gives the lowest gravitational (and baryon) mass and angular momentum.
This results from the fact that an intermediate $\Gamma_\mathrm{th}$ gives the highest torus mass and angular momentum, which are not counted in the merger remnant.
That the torus mass and angular momentum vary with $\Gamma_\mathrm{th}$ in the same way as the mass ejection (as described in Ref.~\cite{Han2025}) reflects a competition between the thermal pressure support, which grows with $\Gamma_\mathrm{th}$, and the gravitational binding effect, the two balancing each other at an intermediate $\Gamma_\mathrm{th}$.

\subsection{Evolution under different loss channels}
\label{sec:transport}

In this subsection, we consider the effect of the different angular momentum loss channels, shown in Fig.~\ref{fig:transport}, where we compare two different EOSs (medium and stiff) with three different $\Gamma_\mathrm{th}$ while the initial total masses of the systems are kept the same.
Seven possible evolution tracks are shown with different color combinations.
By comparing the allowed range of angular momentum at a fixed baryon mass, we find that the stiffer EOS can sustain a larger part of the angular momentum.
The reason is twofold: first, the larger-radius part of the NS carries most of the angular momentum; second, NSs with stiffer EOSs can withstand collapse down to a lower angular momentum, because a stiffer EOS provides stronger pressure support against collapse.
We also notice that the ratio $\mathrm{d}M_\mathrm{b}/\mathrm{d}J$ is largely constant in both the mass-shedding limit and the turning-point line at the higher end of the wedge-shaped plots (where most of the BNS merger remnants of interest sit), with a higher slope for the mass-shedding line.

We start by commenting on the starting position of the remnant.
As shown in Fig.~\ref{fig:stablemap} and discussed in Ref.~\cite{Han2025}, the remnant mass (within a certain density cut) behaves non-monotonically with respect to $\Gamma_\mathrm{th}$, reflecting a competition between the excess of thermal pressure and the gravitational binding effect.
Across the EOSs, we find that a softer EOS gives a higher baryon mass at the same initial gravitational mass and the same thermal index, mainly because the gravitational binding energy is larger for a softer EOS.
For the total angular momentum, however, the trend is not conclusive, since a stiffer EOS retains higher angular momentum in the small and large $\Gamma_\mathrm{th}$ limits, but lower angular momentum at intermediate $\Gamma_\mathrm{th}$.

\begin{figure*}[!t]
  \includegraphics[width=0.98\linewidth]{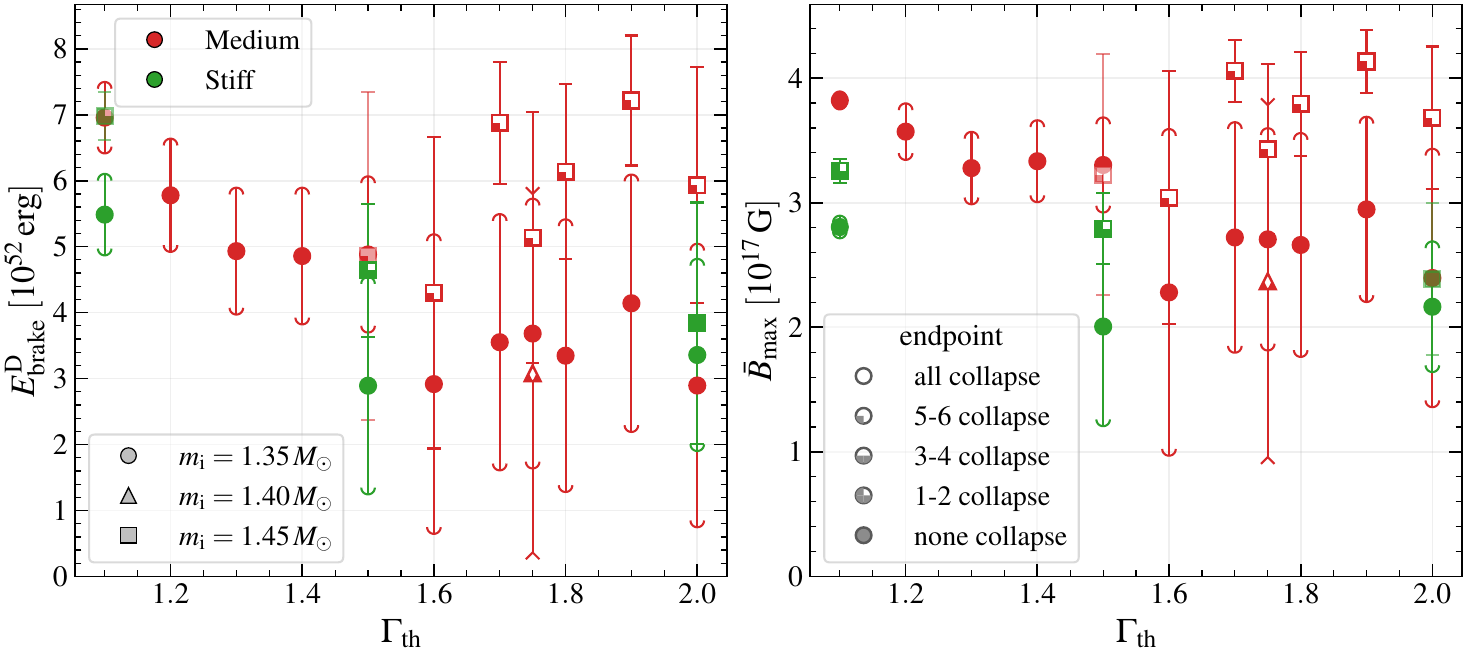}
  \caption{The differential braking energy left at the end stage of the differential rotation, $E^\mathrm{D}_\mathrm{brake}$ (left), and the average magnetic field (with respect to proper volume) that it can generate if all of it goes into the magnetic field, $\bar{B}_{\max}$ (right), against the thermal index.
  Each point is the mean over the counted evolutions of its run, and the error bar is their standard deviation; the color denotes the EOS and the shape the binary mass (the same shape as used in Fig.~\ref{fig:stablemap}).
  Only the medium and the stiff EOS are shown, as the soft one will collapse eventually.
  The cap at either end of an error bar takes the shape of the marker of its point (two arcs for a circle, a Y above the bar and an inverted Y below it for a triangle, a horizontal bar for a square), so that the bars of the three binary masses remain distinguishable where they overlap; a point drawn in a lighter shade (with reduced opacity) is one whose marker falls on top of another.
  The fill of a marker indicates how the evolutions of the run end: empty when all seven of them reach the turning-point line, which we take as collapse, and full when none does, with one quarter of the marker being filled for every two evolutions fewer that collapse.}
  \label{fig:bgamma}
\end{figure*}

We then turn to the slope of the evolution track.
This is mainly the slope $\mathrm{d}M_\mathrm{b}/\mathrm{d}J$, since, as can be directly inferred from Eq.~\eqref{eq:Jloss}, the track is purely horizontal for radiative processes such as neutrino emission and GW emission, whereas the slope of the hybrid part merely reflects the weight of the mass ejection channel in that angular momentum loss combination.
Consequently, a change in the slope reflects the exhaustion of the energy budget of a channel, with its weight reassigned to each of the remaining channels.
In this sense, a track that steepens at late times indicates that the budget for the radiative processes is consumed earlier than that for mass ejection, while the ones that flatten at late times indicate the opposite.

The position of the endpoint results from a combination of the starting point, the slope, and the energy budget.
In every track that includes mass ejection, the baryon mass decreases by $0.1\Msun$, which merely reflects its upper limit.
In particular, the track with more channels ends with a remnant rotating more slowly and having a smaller gravitational mass.
Moreover, the neutrino channel loses the least angular momentum under our assumptions on the energy budgets and the loss ratios.
Thus only the pure neutrino emission track will, in some cases, not reach the mass-shedding limit, so such tracks are not counted in some of the results.
All of these omitted tracks concentrate at the lowest and highest $\Gamma_\mathrm{th}$ (see the two left panels and the lower-right panel of Fig.~\ref{fig:transport}).
The reason differs between the lowest and the highest $\Gamma_\mathrm{th}$, where in the lowest cases it is the large reservoir of angular momentum in the initial stage (see the discussion in Sec.~\ref{sec:remnant}), whereas in the highest ones it is the smaller thermal energy (and hence the smaller amount of angular momentum that the neutrino channel can carry), since a higher $\Gamma_\mathrm{th}$ tends to leave less thermal energy (see the discussion in Sec.~\ref{sec:remnant}).
In general, the remnants with lower initial angular momentum end closer to the turning-point line and rotate more slowly.
We also find that remnants with a stiffer EOS end further toward the mass-shedding limit, which mainly reflects the fact that a stiffer EOS can sustain a smaller angular momentum at the turning-point line, since a larger pressure support for resisting collapse requires less centrifugal effect.

Finally, we turn to the tracks that reach the turning-point line, which we regard as the collapse line without much loss of accuracy~\cite{Gao2026_res}.
The number of tracks with a collapse endpoint is collected in the sixth column of Table~\ref{tab:transport}, which shows that, for higher-mass remnants, collapse is more probable (see the \texttt{M-M29} cases), since the range of angular momentum between the mass-shedding limit and the turning-point line is narrower at a higher baryon mass (see Fig.~\ref{fig:stablemap}).
The probability of collapse becomes smaller for the stiffer EOS (see the \texttt{H-M29} cases), even when the total mass is as high as $2.9\Msun$.
In fact, the probability of collapse remains low in these cases, with only one track leading to collapse at the end even under the largest angular momentum losses assumed here.
For the medium case with $M_\mathrm{tot}=2.9\Msun$, by contrast, the absence of runs with $\Gamma_\mathrm{th}=1.1$--$1.4$ from the table already shows that a low thermal index cannot resist collapse; although a high thermal index resists collapse in the initial post-merger stage, it will not last long if further angular momentum continues to be lost through the three channels considered here.
In each of the seven \texttt{M-M29} runs, six of the seven tracks collapse, and the only surviving track is the pure mass-ejection track, since the remnants already sit close to the turning-point line and only mass loss can allow more angular momentum to be lost while remaining on the stable branch.

\subsection{Energy reservoir and remnant properties at the endpoint}

We investigate the impact of the thermal effect, EOS, and total mass on the extractable energy, magnetic field, and rotation frequency of the remnant in this subsection.
We start by describing the differential braking energy (left panel of Fig.~\ref{fig:bgamma}) that can be extracted from the difference in gravitational mass between the endpoint of the differentially rotating configuration and the uniformly rotating configuration that has the same baryon mass and angular momentum.
The dependence of the result on these factors---thermal index, total mass, and EOS---is very similar for the three quantities $E^\mathrm{D}_\mathrm{brake}$, $\bar{B}_{\max}$, and $f_\mathrm{rem}$.
We focus on each of these factors in the following paragraphs, and later we comment on the trend for each variable that differs from that of the differential braking energy.

\begin{figure*}[!t]
  \includegraphics[width=0.98\linewidth]{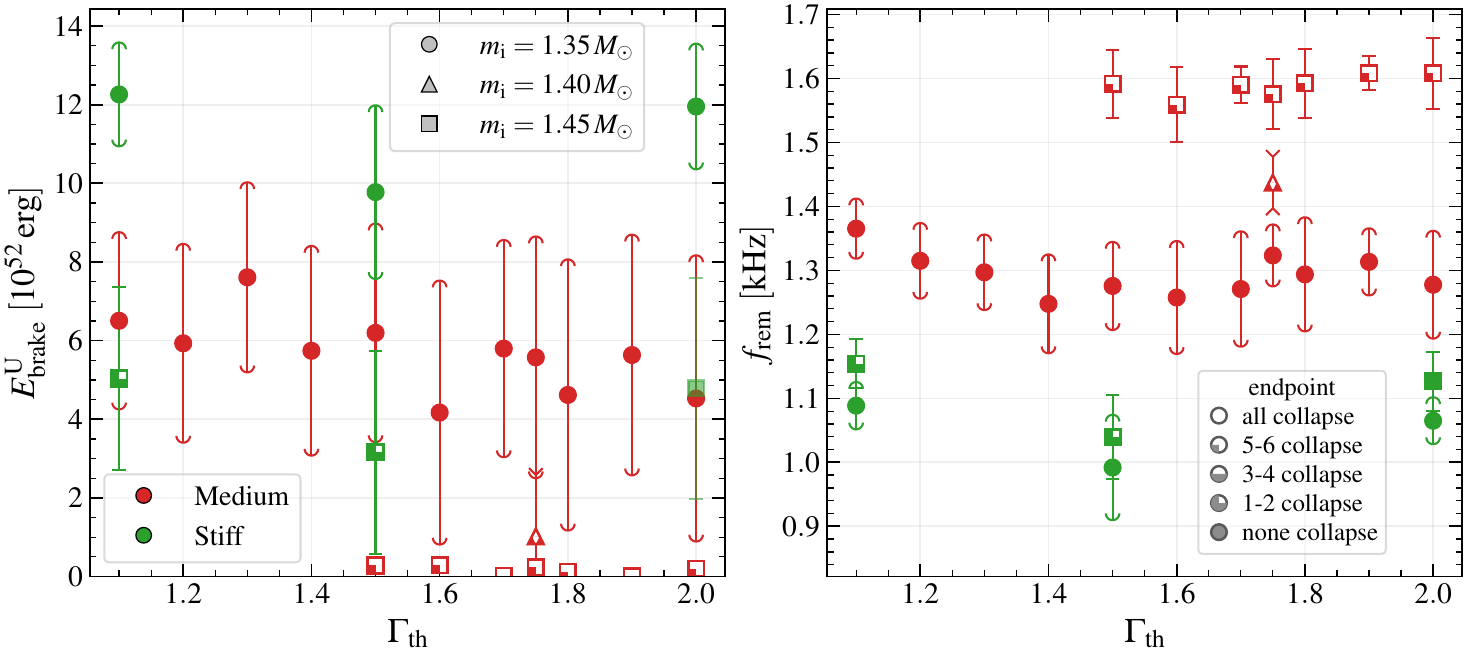}
  \caption{The same as Fig.~\ref{fig:bgamma}, but for the uniform braking energy of the cold equilibrium configuration with the same baryon mass and angular momentum of the remnant at the end of the evolutions, $E^\mathrm{U}_\mathrm{brake}$ of Eq.~\eqref{eq:ext} (left), and its rotation frequency $f_\mathrm{rem}$ (right).}
  \label{fig:trns}
\end{figure*}

We first focus on the thermal effect.
It is observed that the low-$\Gamma_\mathrm{th}$ regime has more differential braking energy, whereas in the high-$\Gamma_\mathrm{th}$ range ($\Gamma_\mathrm{th}\ge1.6$) the value is scattered around $(3$--$5)\times10^{52}~\mathrm{erg}$ for all the non-collapsing scenarios.
This trend can also be seen in the upper-middle panel of Fig.~\ref{fig:stablemap}, where the gravitational mass of the remnant decreases sharply as $\Gamma_\mathrm{th}$ increases, and it scatters around a stable value after a further increase of $\Gamma_\mathrm{th}$.
We therefore attribute the larger differential braking energy to the higher gravitational mass of the remnant at the selected time.
As discussed in Sec.~\ref{sec:remnant}, in the low-$\Gamma_\mathrm{th}$ range a higher $\Gamma_\mathrm{th}$ causes larger mass ejection and a larger torus mass, which leaves the gravitational mass of the remnant smaller as a result of the excess of thermal pressure.
As $\Gamma_\mathrm{th}$ increases, the remnant size is larger and the rotation speed is slower, which causes less loss of matter.
These two effects compete with each other in the high-$\Gamma_\mathrm{th}$ range, which stabilizes the gravitational mass of the remnant and hence the differential braking energy.

As for the EOS effect, we find that in general the stiff EOS cases give smaller differential braking energy for the same initial NS mass; the main reason is not the mass ejection, as its trend with the EOS is not universal (see the discussion in Ref.~\cite{Han2025}).
The main reason is that, as the remnant of the medium EOS is more compact, more of its gravitational binding energy can be released than in the stiff case.

The effect of the total mass is rather simple.
A remnant with a higher initial gravitational mass has a much larger energy reservoir.
Moreover, a higher mass produces more compact objects, so the argument above, which allows the medium EOS case to retain more differential braking energy, still holds.
The only possible exception is the case \texttt{M-M28-G1.75}; in this case, the error bar is rather large, so the position of the mean value alone is not conclusive.

If, in the most extreme case, all the differential braking energy is transferred to the magnetic field, the magnetic field averaged over the whole proper volume is
\begin{equation}
  \bar{B}_{\max}=\sqrt{\frac{8\pi E^\mathrm{D}_\mathrm{brake}}{V_\mathrm{p}}},
  \label{eq:Bmax}
\end{equation}
where $V_\mathrm{p}$ is the proper volume at the endpoint of the uniformly rotating case, defined in Appendix~\ref{app:global_quantities}.
Note that only part of this total EM energy is in the poloidal form, and only this form takes part in the dipole braking which we will discuss in Sec.~\ref{sec:injection}, so we also quote the surface dipole field here,
\begin{equation}
  B_\mathrm{p}=\eta_\mathrm{dip}\bar{B}_{\max},
  \label{eq:Bp}
\end{equation}
with $\eta_\mathrm{dip}$ the fraction of the averaged field that the dipole carries.

The averaged magnetic field (right panel of Fig.~\ref{fig:bgamma}) follows almost exactly the same trend with respect to the thermal index, EOS, and total mass as $E^\mathrm{D}_\mathrm{brake}$; only a few cases show slightly different behavior, which we attribute to their different rotation frequencies $f_\mathrm{rem}$ and hence different proper volumes.
Overall, the differential braking energy spans $(3$--$7)\times10^{52}~\mathrm{erg}$ for most cases considered in this work, which translates into an averaged magnetic field strength of $(2$--$4)\times10^{17}~\mathrm{G}$.
Special attention should be paid here to the higher-mass cases, as many remnants have a high probability of collapsing, and less differential braking energy will be converted into the magnetic field before collapse. Furthermore, the amplified magnetic energy in the \texttt{M-M29} cases might eventually be swallowed by the black hole.

The trend of the uniform braking energy of the equilibrium configuration (left panel of Fig.~\ref{fig:trns}) is very different from that of the differential braking energy, since the two measure different things, $E^\mathrm{U}_\mathrm{brake}$ being what uniform rotation can still release once the differential rotation is gone, which is set by how far the endpoint sits from the reference configuration of its baryon mass.
It is found that the dependence on the thermal index is weak.
In addition, the stiff EOS keeps the largest reservoir, $(1.0$--$1.2)\times10^{53}~\mathrm{erg}$ at $2.7\Msun$.
The reason is twofold: first, it can hold a more slowly rotating configuration of a given baryon mass before reaching the turning-point line, \ie a rather low reference energy; second, it loses less angular momentum than the medium EOS case (see the different length of the evolution tracks in Fig.~\ref{fig:transport}) so that the endpoint is further away from the turning-point line.
Lastly, in contrast to $E^\mathrm{D}_\mathrm{brake}$, $E^\mathrm{U}_\mathrm{brake}$ decreases with total mass instead of increasing, as the heavier remnants end so close to the turning-point line that almost nothing is left, $E^\mathrm{U}_\mathrm{brake}\lesssim0.3\times10^{52}~\mathrm{erg}$ for the medium EOS at $2.9\Msun$, against $(4$--$8)\times10^{52}~\mathrm{erg}$ at $2.7\Msun$.
Appendix~\ref{app:extractable} further locates the endpoints of these evolutions in the cold uniform rotation sequence and evaluates Eq.~\eqref{eq:ext} along with it.

The remnant rotation frequency (right panel of Fig.~\ref{fig:trns}) follows a trend with respect to the thermal-index, EOS, and total-mass effects similar to that of $E^\mathrm{D}_\mathrm{brake}$, and is even more stratified.
Higher-mass remnants end up with higher rotation frequencies, with the medium cases reaching $\sim1.6~\mathrm{kHz}$ and the stiff cases $\sim1.1~\mathrm{kHz}$, for two reasons: first, the initial angular momentum reservoir is higher for higher initial mass (see also Fig.~\ref{fig:stablemap}); second, most of them will collapse and no further spin-down is allowed.
It is also observed that stiffer EOS endpoints rotate more slowly, around $1.0$--$1.2~\mathrm{kHz}$, while the medium cases scatter around $1.2$--$1.4~\mathrm{kHz}$ regardless of the thermal effect.
This merely reflects the fact that the rotation frequency of the remnant at the selected time (which acts as the starting point of the evolution tracks in Fig.~\ref{fig:transport}) is rather low for a stiffer EOS, as the merger occurs at larger radii and with a smaller orbital velocity.
Another way to look at this is that, even if the same order of angular momentum is left in the remnant, the stiffer EOS remnants with larger radii will rotate more slowly.

Finally, our upper limit of EM energy of $(3$--$7)\times10^{52}~\mathrm{erg}$ can be compared with the maximum $E_\mathrm{mag}$ reached in GRMHD simulations: $\sim5\times10^{50}~\mathrm{erg}$ of Ref.~\cite{Neuweiler2025}, $\sim1\times10^{51}~\mathrm{erg}$ of Ref.~\cite{Kiuchi2026b}, $\sim5\times10^{51}~\mathrm{erg}$ of Ref.~\cite{Kiuchi2023}, $\sim7\times10^{51}~\mathrm{erg}$ of Refs.~\cite{Jiang2025, Wen2026}, $\sim1\times10^{52}~\mathrm{erg}$ of Refs.~\cite{Shibata2021a, Rainho2025}, and $\sim3\times10^{52}~\mathrm{erg}$ of Ref.~\cite{Bamber2024b}, as well as the results from GRMHD simulations with the sub-grid terms, where Ref.~\cite{Most2023b} obtained $(\sim1$--$2)\times10^{50}~\mathrm{erg}$, Ref.~\cite{Palenzuela2022} $\sim3\times10^{50}~\mathrm{erg}$, Ref.~\cite{Miret2023} $\sim1\times10^{51}~\mathrm{erg}$, Ref.~\cite{Shibata2021b} $\sim2\times10^{51}~\mathrm{erg}$ and Ref.~\cite{Ng2026b} $\sim4\times10^{51}~\mathrm{erg}$.
We emphasize that there is still room for a few times more magnetic field amplification in the differential braking energy reservoir.
This underlines the need for future high-resolution simulations to reach the true physical bound of magnetic field amplification.
However, our upper limit assumes a highly idealized case in which all the differential braking energy is transferred into the magnetic field, since part of the differential braking energy consists of thermal energy and the kinetic energy of turbulence and oscillations.
The latter may eventually be dissipated into thermal energy and radiated by neutrinos.
Here we aim to set an upper limit on the amplification of the magnetic field driven by differential rotation; channeling part of this differential braking energy into other forms of energy (\ie $E_\mathrm{res}$ in Eq.~\eqref{eq:mass-balance}) would only lower this upper-limit estimate, which we leave to future work.

\subsection{Observational implications}
\label{sec:injection}

Two kinds of energy reservoir are discussed in this work, and their roles and timescales differ.
The first is the differential braking energy defined in Eq.~\eqref{eq:free-energy-final}, which might be largely converted into EM energy; the second is the uniform braking energy of Eq.~\eqref{eq:ext} that the uniformly rotating configuration still holds after the differential rotation is lost.
The uniform braking energy may be injected into the environment quickly, at a rate set by the magnetic field built from the differential braking energy, while the amplified magnetic energy itself will be rearranged and dissipated over a much longer time.
In this subsection we discuss the observational implications of these reservoirs.

Magnetars are invoked to explain various kinds of observations related to GRBs, such as prompt emission, energy injection for the plateaus, X-ray flares, giant flares, quasi-periodic behavior~\cite[see, \eg][]{Usov1992, Dai1998a, Dai1998b, Zhang2001, Gao2006, Dai2006, Fan2006, ZhangD2009, Xiao2018, Zhang2018_GRB_book, Rehman2025, ChenRC2026a, ChenRC2026b}.
Both of the energy reservoirs mentioned above can easily account for the energy of the plateaus that magnetar spin-down is invoked to explain, $10^{49}$--$10^{52}~\mathrm{erg}$~\cite{Rowlinson2013, Xie2022}.
They might also account for the GRB prompt emission energy in some high-energy events if the beaming correction is properly taken into account~\cite{Lv2014}.
Finally, it should be noted that the follow-up of the best-observed BNS merger event, GW170817, did not reveal the existence of a magnetar, and the limits that this places on long-lived remnants are already competitive~\cite{Abbott2018_GW170817_search_remnant, Liu2020, WangH2023}, which can challenge this super-strong magnetar idea.

\subsubsection{Differential braking energy}

One implication, if most of the differential braking energy is converted into magnetic energy and the magnetic field quickly reorganizes mainly into poloidal form (\ie $\eta_\mathrm{dip}\sim\mathcal{O}(1)$), is that the merger remnant can become a very strong electron--positron pair producer.
A very efficient electron--positron pair production process may set in as a result of the large electric field, whose maximum is
\begin{equation}
  \frac{E^\mathrm{el}_{\max}}{E^\mathrm{el}_\mathrm{cr}}=\frac{\eta_\mathrm{dip}\bar{B}_{\max}}{E^\mathrm{el}_\mathrm{cr}}
  \left(\frac{\Omega_\mathrm{rem} R_\mathrm{rem}}{c} \right)^{5/2},
  \label{eq:Emax}
\end{equation}
with $R_\mathrm{rem}$ the equatorial circumferential radius of the endpoint configuration, $\Omega_\mathrm{rem}=2\pi f_\mathrm{rem}$ its angular velocity, and $E^\mathrm{el}_\mathrm{cr}=1.32\times10^{16}~\mathrm{V\,cm^{-1}}$ the critical field of QED~\cite{Usov1992}.
For the endpoints of Table~\ref{tab:transport} it is $(2.7$--$14)\times10^{2}$ times $E^\mathrm{el}_\mathrm{cr}$, \ie the field produced by amplification driven by differential rotation could sit far above the strength at which the vacuum breaks down into electron--positron pairs very efficiently~\cite[see, recent discussions in, \eg][]{Kim2026} and might quickly produce a trapped fireball.
The pairs will later be accelerated to very high energies and, upon release, together with the very strong magnetic field, can power luminous transients~\cite{Xiao2018, Yin2023, Rehman2025}. 
It should also be noted, however, that the post-merger environment is baryon-polluted as a result of the merger ejecta, and that the screening of the pairs still poses a difficulty.

The other topic related to observation is the dissipation of this huge EM energy (which is not very sensitive to $\eta_{\rm dip}$).
Hall drift might become active during the long-term evolution of the merger remnant, with a timescale of
\begin{equation}
\begin{split}
  t_\mathrm{Hall}&=\frac{4\pi n_\mathrm{e} q_\mathrm{e} L^2}{c\,\bar{B}_{\max}}\\
  &=6.4\times10^{2}~\mathrm{yr}
    \left(\frac{L}{1~\mathrm{km}}\right)^{2}
    \left(\frac{10^{17}~\mathrm{G}}{\bar{B}_{\max}}\right),
\end{split}
  \label{eq:thall}
\end{equation}
with $n_\mathrm{e}$ the electron density, $q_\mathrm{e}$ the elementary charge, and $L$ the length over which the field is rearranged, for which we take the thickness of the crust, $L\simeq1$~km.
For $n_\mathrm{e}=10^{36}~\mathrm{cm^{-3}}$, $L=1~\mathrm{km}$ and the fields of the endpoint configurations, this timescale is $(1.5$--$7.5)\times10^{2}$ yr, scaling as $(L/1~\mathrm{km})^2$, and is shown in the last column of Table~\ref{tab:transport}.
Ohmic decay of the same field is far slower, so the drift might dominate in the early evolution.
Both are far longer than the $10^{2}$--$10^{5}$ s of a late-time energy-injection episode (such as an X-ray plateau).
Finally, note that the drift itself just cascades the magnetic field to smaller scales, and that is where ohmic decay, ambipolar diffusion, and eruptive reconnection dissipate the magnetic energy.
The rearrangement of the field will unavoidably further reshape the magnetic structure of the remnant through magnetohydrodynamic instabilities, and may cause fast radio bursts (FRBs)~\cite{Zhang2020, Yang2021, Xiao2022} (host-galaxy analysis also supports the merger-remnant explanation~\cite{WangFY2020}), fast X-ray transients (FXTs)~\cite{Xue2019, Xiao2019}, and other transients on the Hall timescale of the crust, $\sim10^{2}$~yr after the GRB itself.

\subsubsection{Uniform braking energy}

The uniform braking energy may be injected into the environment via the magnetic field on a dipole timescale,
\begin{equation}
  t_\mathrm{dip}=\frac{E^\mathrm{U}_\mathrm{brake}}{L_\mathrm{dip}}
  =\frac{6c^3E^\mathrm{U}_\mathrm{brake}}{B_\mathrm{p}^2R_\mathrm{rem}^6\Omega_\mathrm{rem}^4},
  \label{eq:tdip}
\end{equation}
where $B_\mathrm{p}$ is defined in Eq.~\eqref{eq:Bp}.
For $\eta_\mathrm{dip}=1$ the dipole timescale is very short, as listed in Table~\ref{tab:transport}, far below the $10^{2}$--$10^{5}$ s over which late energy injection is inferred~\cite{Gao2006, Fan2006, Rowlinson2013}, but it scales as $\eta_\mathrm{dip}^{-2}$, so a smaller value of this factor can relieve the tension.
Since the result depends significantly on $\eta_\mathrm{dip}$, below we discuss two extreme cases for it.

Two limiting values of $\eta_\mathrm{dip}$ are possible, \ie $\eta_\mathrm{dip}\sim\mathcal{O}(1)$ and $\eta_\mathrm{dip}\ll1$.
Both of these two extreme cases can occur, as in current GRMHD simulations the ratio of the toroidal to poloidal magnetic energy does not converge to a stable value, and different amplification processes may imply these two extreme cases: $\eta_\mathrm{dip}\sim\mathcal{O}(1)$ as the dynamo process might predict and $\eta_\mathrm{dip}\ll1$ as the winding process alone would indicate.

In the first limit ($\eta_\mathrm{dip}\sim\mathcal{O}(1)$), a fast generation of electron--positron pairs can set in, with the rotation energy quickly injected into the environment within a few tens of milliseconds and possibly producing a very energetic GRB, while the later dissipation process and GW amplitude might be much weaker than in the second case (which we discuss below) or even absent.
The remnant might then quickly collapse into a black hole or spin down to a very slowly rotating hot NS.
These outcomes depend strongly on the timescale and on the final equilibrium ratio between the toroidal and poloidal magnetic energy in the post-merger remnant, calling urgently for further high-resolution simulations.

In the second limit ($\eta_\mathrm{dip}\ll1$), the dipole timescale can be significantly longer, which can explain the timescale of the plateau phase; the initial electron--positron pair production is reduced, and the jet it generates (if any) will be less energetic.
However, with this huge magnetic energy stored in the toroidal field inside the NS, Hall drift will later cascade it to smaller scales and trigger a series of significant energy releases.
So we may have a weak jet and a low energy injection for the plateau, while a luminous FXT or FRB appears in the later field-dissipation stage.
In this case, the dominant toroidal field can also slowly convert into the poloidal one, and might lead to time-varying braking indices~\cite{Sarin2020, Xie2022}.
Additionally, since further spin-down of the uniform rotation by the magnetic field alone is not significant while a large amount of rotation energy still persists, other spin-down processes may act, such as GW emission.
Three main processes can cause further GW spin-down of the remnant.
The first is continuous GW emission as a result of magnetically induced ellipticity~\cite[for observational implications, see, \eg][]{Fan2013, Xie2022}.
A strong initial toroidal magnetic field together with a special cooling history can even lead to a further inclination-angle turning and enhance GW emission~\cite{DallOsso2008, WangJS2017}, and might contribute to wobbling behavior of the jet~\cite{WangH2026}.
The second is GW emission caused by the $T/W$ instability~\cite{Lai1995, Morsink1998, Watts2005, Friedman-Nikolaos2013, Muhlberger2014, Paschalidis2016b, Gao2026_res}.
Table~\ref{tab:transport} lists the ratio of the rotational energy to the binding energy of the configuration each track ends at, $T_\mathrm{kin}/E_\mathrm{bind}=0.08$--$0.14$.
For a uniformly rotating star the $m=2$ bar mode is secularly unstable if $T_\mathrm{kin}/E_\mathrm{bind} > 0.115-0.048M/M_\mathrm{TOV}$~\cite{Morsink1998}.
Extrapolated to our supramassive endpoints, this empirical relation places the vast majority of the remnants above this threshold, so the secular instability may operate and radiate GWs for the remnants that reach uniform rotation in this work.
The third channel of further angular momentum loss by GW for the uniformly rotating remnant is the $r$-mode instability; if it operates, it radiates at $4/3$ of the spin frequency, $1.3$--$2.1~\mathrm{kHz}$, for the remnants shown in this work.
However, this channel depends on the cooling history of the remnant, and a detailed analysis is beyond the scope of this work.

\section{Discussion}
\label{sec:discussion}

In this work, we reuse previous numerical simulation results, choosing those that survive $25~\mathrm{ms}$ after the merger in general-relativistic hydrodynamic simulations, and follow the merger remnant along seven possible angular momentum loss tracks.
We then compare the final possible evolution endpoint with the uniformly rotating configuration and find an energy reservoir of $(3$--$7)\times10^{52}~\mathrm{erg}$.
This sets an upper limit on the magnetic field amplification driven by differential rotation, such as winding, MRI, and TSI.
The limit is $(2$--$4)\times10^{17}~\mathrm{G}$ for the volume-averaged magnetic field; the actual value will be lower because part of the free energy is thermalized and radiated away.
We note that a magnetic field of this magnitude in the remnant is consistent with previous estimates that the BNS merger remnant may become a magnetar, as well as with recent high-resolution simulations~\cite{Desai2026, Kiuchi2026a, Kiuchi2026b}.
This upper-limit estimate shows that there is still ample room for magnetic field amplification beyond the current saturation level, and further high-resolution numerical simulations are needed in the future to find out the true upper bound.

We clearly separate the energy reservoir in the post-merger remnant into two parts: the braking energy stored in the differential rotation down to the uniformly rotating configuration, and the braking energy that the uniform rotation still holds once the differential rotation is lost.
The former relates to the ability to amplify the magnetic field, and the latter to the energy that can be released through the spin-down of the uniformly rotating remnant through the amplified magnetic field and other channels.
The former is given by the stored braking energy of the differential rotation, $(3$--$7)\times10^{52}~\mathrm{erg}$; the latter is of the same order, from almost nothing for the heaviest remnants, which end on the turning-point line with no rotation energy left, up to $1.2\times10^{53}~\mathrm{erg}$ for the stiff EOS at the lighter mass.
The differential braking energy that is channeled into the magnetic field will dissipate on a timescale far longer than years and might power repeating FRBs, X-ray bursts, and other phenomena in the long term through the topological change and dissipation of the amplified magnetic field.
The braking energy of the uniform rotation, on the other hand, might be quickly dissipated by the super-strong magnetic field, or later by GW emission driven by the deformation of the remnant or by secular or $r$-mode instabilities.
We note that each of these reservoirs can easily supply a plateau phase of the GRB, except in the collapsing case.
The strong amplification of magnetic energy with a field strength far above the critical QED limit and the fast dissipation of the huge reservoir of uniform rotation may even lead to GRB prompt emission.
All these aspects need further investigation.
Among them, four key questions need to be clarified in the future and would greatly help us understand this topic:
(1) how many BNS merger remnants are non-collapsing, which depends strongly on the EOS and on further study of the angular momentum transfer mechanism;
(2) the true conversion efficiency from the free energy into the magnetic field, \ie the true upper limit of the amplified magnetic energy;
(3) the mechanism and timescale by which the dominant toroidal magnetic field is transferred into the poloidal one (\ie the time evolution of $\eta_\mathrm{dip}$) and the final equilibrium ratio between the two;
(4) the electrodynamics of supercritical magnetic fields in the complex environment of the post-merger surrounding matter.

This work has three main limitations, which future observations and simulations can help to remove.
(i) The parameter space: the total mass and mass ratios scanned in this work are very limited as a result of computational resource limits, and thus the conclusions are limited to a very small parameter space.
Regarding EOS uncertainty, it is well known that the EOS remains loosely constrained with current best theoretical and observational knowledge~\cite[see, \eg][]{Chatziioannou2020, Jiang2019, Han2022, Jiang2022, Essick2023, Zhou2023, Huang2026b, Tang2025b}; although we considered three typical EOSs that fulfill the current limits and span the allowed range of $\Lambda_{1.4}$ evenly in this work, the soft one will eventually collapse even at the lightest total mass and largest thermal index considered here, which further prevents us from drawing stronger conclusions.
(ii) Regarding the angular momentum transfer, the loss ratio and the weight of the different angular momentum loss channels are assumed constant until one channel reaches its upper limit of the budget, which is highly idealized and needs further improvement; however, the evolution of the post-merger remnant is highly nonlinear, and perhaps only high-resolution simulations can further enhance our understanding of this topic.
(iii) Regarding the conversion efficiency of the differential braking energy into the magnetic field, in this work we assume the maximum conversion efficiency, and thus our results only serve as an upper limit for magnetic field amplification driven by differential rotation during the late post-merger remnant evolution; the real conversion efficiency and the true upper limit of magnetic field amplification remain to be determined by future simulations.

\begin{acknowledgments}
The authors thank S. P. Tang in particular for stimulating discussions and for providing the data for comparison, as well as H. Wang for useful discussions and Prof. Y. Z. Fan for useful comments on the manuscript.
We acknowledge support from the National Natural Science Foundation of China under Grants No.~12603080 and No.~12233011, and from the Project for Young Scientists in Basic Research (Grant No.~YSBR-088) of the Chinese Academy of Sciences.
We also acknowledge \texttt{DeepSeek-V4.1-Flash} for verifying the formulas, polishing the figures, and improving the grammar.

\end{acknowledgments}

\appendix

\begin{figure*}[t]
\centering
\includegraphics[width=0.98\linewidth, height=0.45\linewidth]{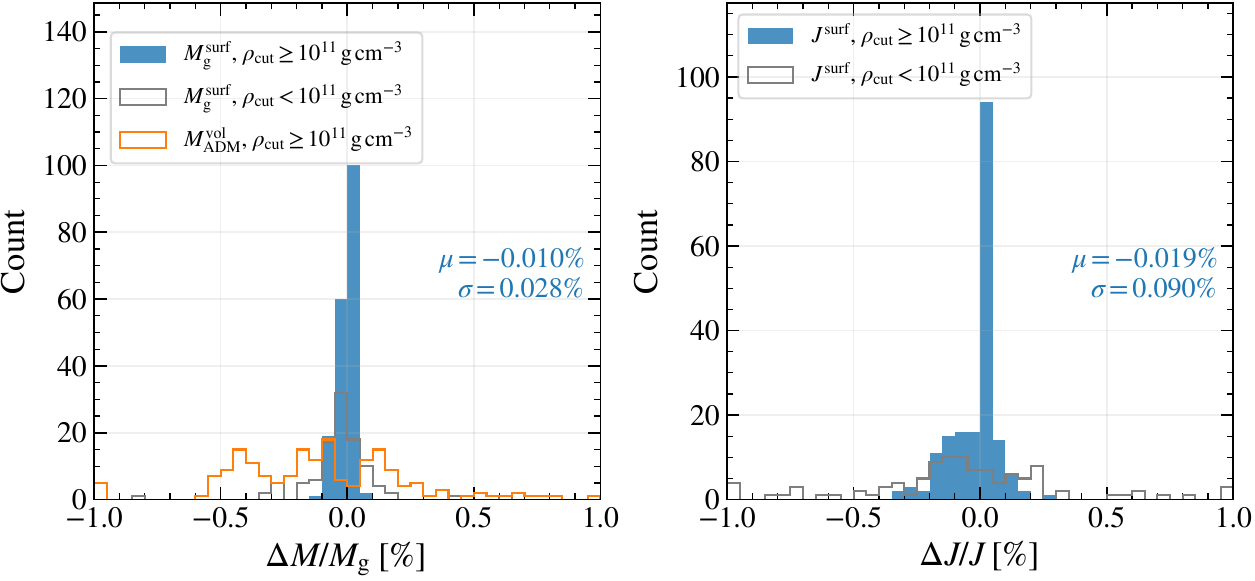}
\caption{Relative differences between the definitions of $M_\mathrm{g}$ and $J$ at $t_\mathrm{s}$.
The reference Komar mass (left) and angular momentum (right) are defined in Eqs.~\eqref{eq:app_Mg} and~\eqref{eq:app_J}, respectively.
Each of the 26 simulations contributes one entry per cut: 182 entries for the seven cuts with $\rho_\mathrm{cut}\geq10^{11}\,\mathrm{g\,cm^{-3}}$ and 104 for the four lower cuts.
Left: difference relative to the surface form of the Komar mass (see Eq.~\eqref{eq:app_Mgsurf}), for the high (blue) and low (gray) density cuts; the orange bins show the difference relative to the ADM volume integral.
Right: the same comparison for $J$ with $J_{\rm surf}$ defined in Eq.~\eqref{eq:app_Jsurf}.}
\label{fig:komar}
\end{figure*}

\section{Global quantities of the remnant}
\label{app:global_quantities}

To compare the merger remnant with the uniformly rotating equilibria of the \rns code, we evaluate its global quantities inside the isodensity surface $\partial V$ on which $\rho=\rho_\mathrm{cut}$.
All integrals extend over the region $V$ enclosed by $\partial V$, in which $\rho\geq\rho_\mathrm{cut}$.
In this appendix we define these quantities, show that they are gauge independent, and check how well this holds at the selected time $t_\mathrm{s}$.

\subsection{Definitions}

We write the metric in the $3+1$ form with lapse $\alpha$, shift $\beta^i$, and spatial metric $\gamma_{ij}$, as used in \texttt{SACRA-MPI}.
For the metric of Eq.~\eqref{eq:metric} used in \rns, these are
\begin{equation}
\alpha=N,\qquad \beta^\varphi=-\omega,\qquad \sqrt{\gamma}=A^2Br^2\sin\theta .
\label{eq:app_3p1}
\end{equation}
The baryon mass, the thermal energy, and the proper volume are
\begin{align}
M_\mathrm{b}&=\int_V\rho_*\,\mathrm{d}^3x, \label{eq:app_Mb}\\
E_\mathrm{th}&=\int_V\rho_*\big[\varepsilon-\varepsilon_\mathrm{cold}(\rho)\big]\,\mathrm{d}^3x, \label{eq:app_Eth}\\
V_\mathrm{p}&=\int_V\sqrt{\gamma}\,\mathrm{d}^3x, \label{eq:app_Vp}
\end{align}
where $\rho_*=\rho\,\Gamma\sqrt{\gamma}$ is the conserved rest-mass density.
The thermal energy vanishes for the cold equilibria.
For a stationary circular flow, $M_\mathrm{b}$ and $E_\mathrm{th}$ do not depend on the slicing, whereas $V_\mathrm{p}$ refers to the $t=\mathrm{const}$ slices of the stationary metric.

The gravitational mass and the angular momentum are the Komar integrals used in analyzing the results from both codes, in their volume form~\cite{CST94, Stergioulas1995, Gourgoulhon2010},
\begin{align}
M_\mathrm{g}&=\int_V\big(-2T^t{}_t+T^\mu{}_\mu\big)\sqrt{-g}\,\mathrm{d}^3x \nonumber\\
&=\int_V\big[\alpha(E+S)-2\beta^iS_i\big]\sqrt{\gamma}\,\mathrm{d}^3x, \label{eq:app_Mg}\\
J&=\int_V T^t{}_\varphi\sqrt{-g}\,\mathrm{d}^3x=\int_V S_i\varphi^i\sqrt{\gamma}\,\mathrm{d}^3x, \label{eq:app_J}
\end{align}
where $E$, $S_i$, and $S$ are the energy density, the momentum density, and the trace of the stress tensor measured by the Eulerian observer, and $\varphi^i$ is the axial vector, with the axis passing through the center of mass of the remnant.
The rotational kinetic energy in \rns is defined in a similar way,
\begin{equation}
T_\mathrm{kin}=\frac12\int_V\Omega\,T^t{}_\varphi\sqrt{-g}\,\mathrm{d}^3x ,
\label{eq:app_T}
\end{equation}
with $T^t{}_\varphi=(e+p)\,j$ for a perfect fluid, where $j$ and $\Omega$ are the same as in Eq.~\eqref{eq:rot_prof}.
For the uniformly rotating equilibria it reduces to
\begin{equation}
T_\mathrm{kin}=\frac12\,\Omega J .
\label{eq:app_Trns}
\end{equation}
The gravitational binding energy $E_\mathrm{bind}$ that enters the ratio $T_\mathrm{kin}/E_\mathrm{bind}$ of rotational instabilities follows from the decomposition of the gravitational mass~\cite{CST94},
\begin{equation}
E_\mathrm{bind}=M_\mathrm{p}+T_\mathrm{kin}-M_\mathrm{g},\qquad M_\mathrm{p}=\int_V e\,\Gamma\sqrt{\gamma}\,\mathrm{d}^3x ,
\label{eq:app_EB}
\end{equation}
where the proper mass $M_\mathrm{p}$ adds the internal energy to the baryon mass, $e\,\Gamma\sqrt{\gamma}=\rho_*(1+\varepsilon)$.
Note that, by this definition, the binding energy is always positive for a bound configuration.

\begin{figure*}[t]
  \centering
  \includegraphics[width=0.98\linewidth, height=0.5\linewidth]{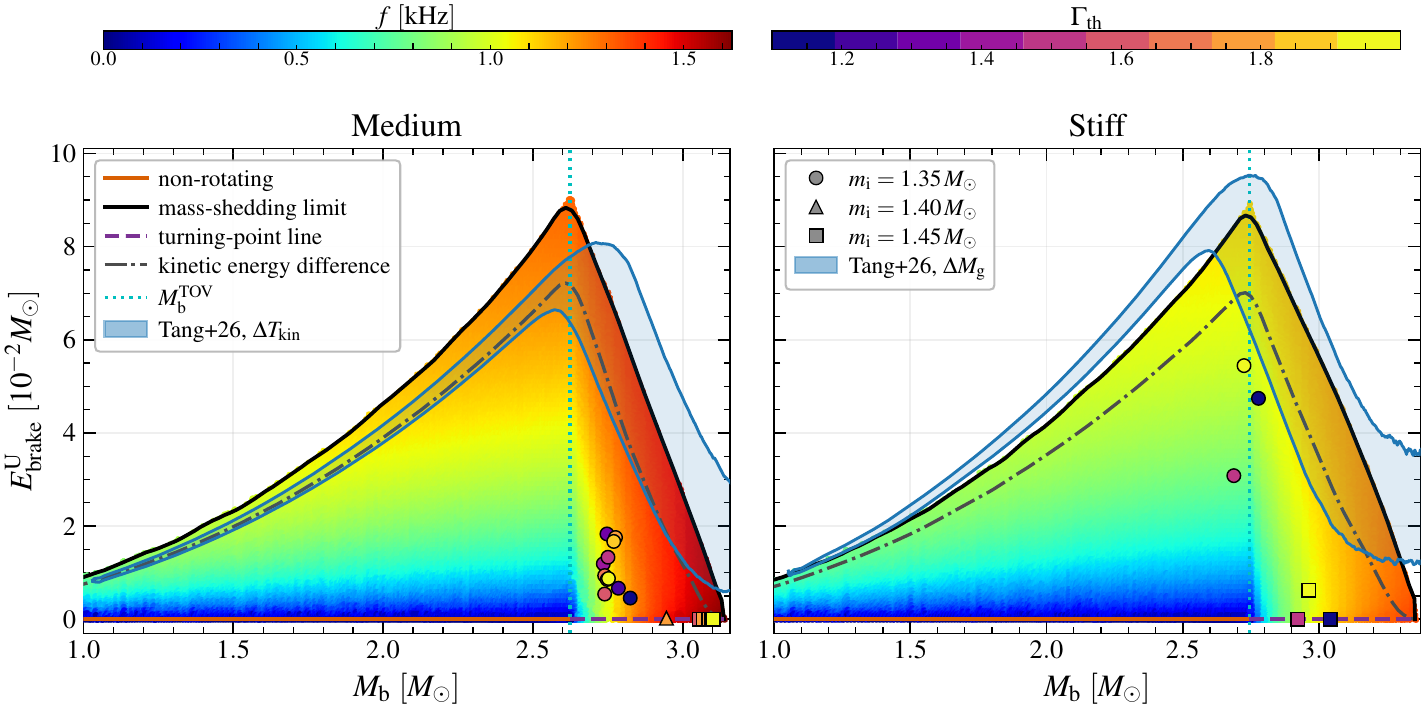}
  \caption{The cold uniformly rotating sequence of the medium (left) and stiff (right) EOS in the baryon mass--uniform braking energy plane, each equilibrium of the \rns sequence being one point, colored by its rotation frequency (color bar in the upper left).
The lower edge of the band is the reference from which the energy is measured: the non-rotating sequence (solid orange) up to its maximum mass (cyan, dotted vertical line) and the turning-point line (violet dashed), the smallest angular momentum configuration to which a remnant of that baryon mass can spin down.
The upper edge is the mass-shedding limit (solid black); the dash-dotted gray curve is the same edge for the kinetic energy difference alone, \ie $T_\mathrm{kin}(f)-T_\mathrm{kin}^\mathrm{ref}$.
The points are the endpoints of the three-channel hybrid evolution, iii-GNM, of every run, placed at the uniform model of the same angular momentum and baryon mass; their shape gives the binary mass and their color the thermal index (color bar in the upper right), the same as in Fig.~\ref{fig:stablemap}.
The bands of lighter color overlaid on the two panels are the same energy obtained by Ref.~\cite{Tang2025a} for its GP ensemble of EOSs, taken from its public data release~\cite{Tang2026data}: the kinetic-energy difference on the left and the difference of gravitational masses at fixed baryon mass on the right.
Each of them is the 68\% credible interval of the EOSs sampled with the posterior weights of that data release.}
  \label{fig:extractable}
\end{figure*}

\subsection{Gauge independence}

Obviously the quantities we compare between the two codes \rns and \texttt{SACRA-MPI} must not depend on the gauge choices.
The volume forms follow from the Komar surface integrals over $\partial V$ through the Einstein equations.
Taking $\partial_t$ as the timelike Killing vector, the surface forms read
\begin{align}
M_\mathrm{g}^\mathrm{surf}&=\frac{1}{4\pi}\oint_{\partial V}\big(D^i\alpha-K^i{}_j\beta^j\big)\,\mathrm{d}S_i, \label{eq:app_Mgsurf}\\
J^\mathrm{surf}&=\frac{1}{8\pi}\oint_{\partial V} K^i{}_j\varphi^j\,\mathrm{d}S_i, \label{eq:app_Jsurf}
\end{align}
where $K_{ij}$ is the extrinsic curvature and $D_i$ the covariant derivative of $\gamma_{ij}$.
In a stationary and axisymmetric spacetime with circular flow (conditions we test quantitatively in the next subsection), the world tube swept by $\partial V$ is generated by the two Killing vectors, so that no Komar flux crosses it.
The integrals then depend only on $\partial V$, which is defined by the scalar $\rho$, and on the normalization of the timelike Killing vector at infinity.
They depend neither on the slicing nor on the spatial coordinates.
ADM-type integrals are not used as they lack this property on a finite surface, and their volume form~\cite{Shibata2005} contains terms of the conformal metric that do not vanish in the $\Gamma$-driver gauge.
This is the main reason that we do not use the ADM mass in comparing with the \rns results and that the comparison of ADM mass with the Komar one is merely a reference, and cannot be interpreted as systematic error.

\subsection{Conditions for stationarity and axisymmetry}

The gauge independence above relies on the stationary and axisymmetric assumptions; here we show how much the merger remnant deviates from them at the selected time.
If we apply the Komar-type integrals for mass and angular momentum in a non-stationary and non-axisymmetric spacetime, the surface and volume forms differ by
\begin{align}
M_\mathrm{g}^\mathrm{surf}-M_\mathrm{g}&=-\frac{1}{4\pi}\int_V\Big(\partial_tK+\frac12K^{ij}\partial_t\gamma_{ij}\Big)\sqrt{\gamma}\,\mathrm{d}^3x, \label{eq:app_dMg}\\
J^\mathrm{surf}-J&=\frac{1}{8\pi}\int_V\big(K^{ij}D_{(i}\varphi_{j)}+\varphi^iD_iK\big)\sqrt{\gamma}\,\mathrm{d}^3x. \label{eq:app_dJ}
\end{align}
The difference in mass measures the non-stationarity of the metric in the simulation gauge, and the difference in angular momentum measures its non-axisymmetry.
At $t_\mathrm{s}$ of Eq.~\eqref{eq:ts}, the $m=2$ deformation traced by $h_{22}$ has decayed below about $5\%$ of its peak, so that the remnant evolves approximately secularly, \ie the deviations above are very small.

Figure~\ref{fig:komar} confirms this for the 26 simulations selected in this work.
For $\rho_\mathrm{cut}\geq10^{11}\,\mathrm{g\,cm^{-3}}$ (seven density cuts in total), the two forms agree to $-0.010\%\pm0.028\%$ (all within $0.11\%$) for $M_\mathrm{g}$ and to $-0.019\%\pm0.090\%$ (all within $0.33\%$) for $J$.
Lower density cuts reach the non-stationary outer disk and the outflow at $80$--$800~\mathrm{km}$, and the scatter grows to $0.15\%$ and $0.51\%$.
The ADM volume integral differs from $M_\mathrm{g}$ by $-0.11\%\pm0.36\%$ and serves merely as a reference, since ADM integrals within a finite volume are not gauge independent.
We therefore use the volume forms of Eqs.~\eqref{eq:app_Mg} and~\eqref{eq:app_J} with $\rho_\mathrm{cut}=10^{11}\,\mathrm{g\,cm^{-3}}$, which enclose the remnant and have a well-defined gravitational mass and angular momentum that can be directly compared with the ones calculated in \rns.

\subsection{Torus}
\label{append1:torus}

Following Ref.~\cite{Fujibayashi2017}, the matter outside $V$ that remains bound and has a density above $10^{7}\,\mathrm{g\,cm^{-3}}$ is described as a torus,
\begin{equation}
V_\mathrm{torus}=\big\{10^{7}\,\mathrm{g\,cm^{-3}}\leq\rho<\rho_\mathrm{cut},\ -u_t<1\big\},
\label{eq:app_Vtorus}
\end{equation}
where the ejecta are identified by the geodesic criterion $-u_t\geq1$ and excluded.
$M_\mathrm{b,torus}$ and $J_\mathrm{torus}$ of Eq.~\eqref{eq:Jdef} are the integrals of Eqs.~\eqref{eq:app_Mb} and~\eqref{eq:app_J} over $V_\mathrm{torus}$, and $M_\mathrm{g}(t_\mathrm{s})$ is given by Eq.~\eqref{eq:app_Mg}.
As a final note, for $\rho_\mathrm{cut}=10^{11}\,\mathrm{g\,cm^{-3}}$ the excluded dynamical ejecta amount to at most $11\%$ of the torus mass, and including them would change $R_\mathrm{torus}$ by at most $1.1\%$.

\section{Maximum uniform braking energy}
\label{app:extractable}

\begin{figure}[!t]
  \centering
  \includegraphics[width=\columnwidth]{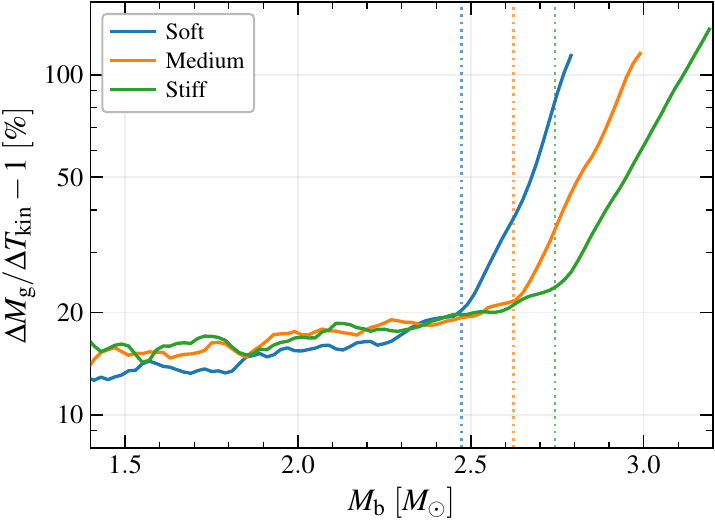}
  \caption{The excess of the gravitational mass difference over the kinetic energy difference at a fixed baryon mass, $(\Delta M_\mathrm{g}/\Delta T_\mathrm{kin}-1)$, along the mass-shedding sequence of the three cold EOSs.
Both differences are measured from the reference configuration of the same baryon mass, as in Fig.~\ref{fig:extractable}, and the dotted vertical lines mark the baryon mass of the TOV configuration of each EOS.}
  \label{fig:mgspr}
\end{figure}

In the main text we studied the braking energy stored in the differential rotation (Sec.~\ref{sec:budget}), whereas in this appendix we evaluate the braking energy of the uniform rotation, Eq.~\eqref{eq:ext} of Sec.~\ref{sec:uniform}, along the cold equilibrium sequences.

Figure~\ref{fig:extractable} shows the whole cold branch of the medium and stiff EOSs in the $(M_\mathrm{b},E^\mathrm{U}_\mathrm{brake})$ plane, colored by rotation frequency $f$.
Up to the maximum baryon mass of the non-rotating sequence, $M_\mathrm{b}^\mathrm{TOV}=2.62$ and $2.74\Msun$, the reference is the non-rotating configuration itself, the one of zero angular momentum; beyond it the reference is the configuration at the turning-point line of that $M_\mathrm{b}$.
The band peaks at $8.8$ and $8.7\times10^{-2}\Msun$, \ie $1.6$ and $1.5\times10^{53}~\mathrm{erg}$, at $M_\mathrm{b}=2.62$ and $2.74\Msun$, $\sim 2.2$ times the largest differential braking energy among the medium and stiff EOS in Table~\ref{tab:transport}.

The points shown are the endpoints of the three-channel hybrid evolution, iii-GNM, of every run, placed at the uniform model of the same $(J,M_\mathrm{b})$.
For the heavier binaries the endpoint falls on the lower edge, $E^\mathrm{U}_\mathrm{brake}\lesssim0.6\times10^{-2}\Msun$, as the three channels have carried away most of the energy that uniform rotation could still release, and the remnant is near the turning-point line,
whereas the lighter ones stop short of it, at $(0.5$--$1.8)\times10^{-2}\Msun$ for the medium EOS and $(3.1$--$5.4)\times10^{-2}\Msun$ for the stiff one.

We then compare the results with those found in Ref.~\cite{Tang2025a}; the two methods of counting the extractable energy are shown in Fig.~\ref{fig:extractable}, \ie the kinetic energy difference on the left (which directly corresponds to their Fig.~3) and the gravitational mass difference on the right.
Our values are slightly higher than those found in Fig.~3 of Ref.~\cite{Tang2025a}, as they calculated the free energy using the kinetic energy difference; however, the comparison is not conclusive, since they used an ensemble of EOSs (while our single panel in Fig.~\ref{fig:extractable} corresponds to one EOS only), and their ensemble is the more general case.
Comparing the uniform braking energy obtained from the gravitational mass (solid black lines) with that obtained from the kinetic energy difference (dash-dotted gray lines) for the mass-shedding sequence, we find (see also Fig.~\ref{fig:mgspr}) a difference of $\sim 10\%$ to $\sim 20\%$ between these two methods below $M_{\mathrm{b}}^{\mathrm{TOV}}$, and this difference roughly increases with $M_{\mathrm{b}}$.
Above $M_{\mathrm{b}}^{\mathrm{TOV}}$, however, the difference grows quickly and can exceed $100\%$ where the extractable energy itself drops steeply.
These differences are consistent with those found when comparing the two different datasets from the public data release~\cite{Tang2026data} of Ref.~\cite{Tang2025a}, and highlight the importance of using the gravitational mass difference to calculate the extractable energy in rotating NSs with $M_{\mathrm{b}}>M_{\mathrm{b}}^{\mathrm{TOV}}$, where most BNS merger remnants sit.

\bibliography{goodbib}

\end{document}